\documentclass[letterpaper,english,reprint,pre,longbibliography,aps]{revtex4-1}
\usepackage[T1]{fontenc}
\usepackage[utf8]{inputenc}
\usepackage{babel}
\usepackage{color}
\usepackage{amsmath}
\usepackage{amssymb}
\usepackage{mathtools}
\usepackage{bm}
\usepackage[unicode=true,
 bookmarks=true,bookmarksnumbered=false,bookmarksopen=false,
 breaklinks=false,pdfborder={0 0 1},backref=false,colorlinks=false]
 {hyperref}

\makeatletter

\pdfpageheight\paperheight
\pdfpagewidth\paperwidth

\newcommand{\tD}{\tau_D}

\newcommand{\tDL}{\tilde{t}}
\newcommand{\xDL}{\tilde{x}}
\newcommand{\epsDL}{\tilde{\epsilon}}
\newcommand{\PDL}{\tilde{P}}
\newcommand{\aDL}{\tilde{a}}
\newcommand{\DDL}{\tilde{D}}
\newcommand{\jDL}{\tilde{j}}

\newcommand{\An}{\tilde{\mathcal{A}}_n}
\newcommand{\Aindex}[1]{\tilde{\mathcal{A}}_{#1}}
\newcommand{\Dn}{\tilde{\mathcal{D}}_n}
\newcommand{\Dindex}[1]{\tilde{\mathcal{D}}_{#1}}
\newcommand{\Dzero}{\tilde{\mathcal{D}}_0}
\newcommand{\Qindex}[1]{\tilde{\mathcal{Q}}_{#1}}

\newcommand{\xDLalpha}{\xDL_{\alpha}}

\begin{document}

\title{Short-time Fokker-Planck propagator beyond the Gaussian approximation}
\author{Julian Kappler}
\email{jkappler@posteo.de}

\affiliation{Arnold Sommerfeld Center for Theoretical Physics (ASC), Department of Physics, 
Ludwig-Maximilians-Universit\"at
 M\"unchen, Theresienstra{\ss}e 37, D-80333 Munich, Germany}
\date{\today}
\begin{abstract}
We present a
 perturbation approach to calculate the short-time propagator, or transition density, 
 of the one-dimensional Fokker-Planck equation,
to in principle arbitrary order in the time increment.
Our approach preserves probability exactly and allows us to evaluate expectation values
of analytical observables to in principle arbitrary accuracy;
to showcase this, we derive perturbation expansions 
for the moments of the spatial increment, 
the finite-time Kramers-Moyal coefficients,
and the mean medium entropy production rate.
For an explicit multiplicative-noise system with available analytical solution, 
we validate all our perturbative results.
Throughout, we compare our perturbative results to those obtained from the
 widely used Gaussian approximation of the short-time propagator;
we demonstrate that this Gaussian propagator leads to errors that can be 
many orders of magnitude
larger than those resulting from our perturbation approach. 
Potential applications of our results include parametrizing diffusive stochastic dynamics from observed time series,
and sampling path spaces of stochastic trajectories numerically.
 % is only accurate
\end{abstract}
\maketitle

\section{Introduction}

Diffusive stochastic processes are ubiquitous 
in several branches of science, 
including physics, chemistry, and biology \cite{risken_fokker-planck_1996,
van_kampen_stochastic_2007,
gardiner_stochastic_2009,
bressloff_stochastic_2014}.
The time evolution of a reaction coordinate subject to diffusive stochastic dynamics 
can be described by a stochastic differential equation (SDE), which 
in the physics literature is usually called
the overdamped Langevin equation, or, % \textcolor{red}{[cite]},
equivalently, 
or by its
associated Fokker-Planck equation (FPE) \cite{risken_fokker-planck_1996,
gardiner_stochastic_2009}. % for the spatial distribution of 
While the former constitutes a description on the level 
of stochastic realizations of the reaction coordinate,
the latter describes the time evolution of the probability density for observing
a the reaction coordinate at a given location.
Of particular interest here is 
the short-time propagator, 
i.e.~the distribution of the spatial increment $\Delta x$ after a short time increment
$\Delta t$, for a particle starting  at a given initial position $x_0$.
This propagator allows to calculate expectation values, which in turn
 can be used to parametrize the FPE from observed time 
series \cite{gladrow_experimental_2019,
gladrow_experimental_2021,
thorneywork_resolution_2024}.
Furthermore, the short-time propagator is a starting point to derive the path-integral
representation of the stochastic process via time-slicing \cite{haken_generalized_1976,
graham_path_1977,
hunt_path_1981,
adib_stochastic_2008,
arnold_langevin_2000,
seifert_stochastic_2012,
tang_summing_2014,
chow_path_2015,
weber_master_2017,
cugliandolo_rules_2017,
cugliandolo_building_2019,
de_pirey_path_2023}.
In its discretized form, the path integral 
can be used for Bayesian parameter inference \cite{dacunha-castelle_estimation_1986,
gilks_markov_1995,
elerian_note_1998,
tse_estimation_2004,
beskos_mcmc_2008,
lin_generating_2010,
durmus_fast_2016,
bera_fast_2017,
pieschner_bayesian_2020}
or sampling of transition ensembles \cite{fujisaki_onsagermachlup_2010,
fujisaki_multiscale_2013}
 via Markov Chain Monte Carlo (MCMC);
in its continuum form, it has been used 
to to analyze transitions between metastable states \cite{dykman_theory_1979,
dykman_optimal_1992,
adib_stochastic_2008}
or to quantify irreversibility \cite{maes_time-reversal_2003,
seifert_entropy_2005,
seifert_stochastic_2012,
cates_stochastic_2022}.

The most widely used approximate expression for the short-time propagator uses
 a Gaussian distribution for the spatial increment $\Delta x$, and is based
  on an Euler-Maruyama discretization
of the underlying SDE \cite{hunt_path_1981,
arnold_langevin_2000,
adib_stochastic_2008,
seifert_stochastic_2012,
tang_summing_2014,
chow_path_2015,
cugliandolo_rules_2017,
cugliandolo_building_2019,
moreno_conditional_2019,
de_pirey_path_2023}.
There seem to exist few explicitly known approximate 
short-time propagators beyond the Gaussian approximation in the literature.
Elerian \cite{elerian_note_1998} 
provides a closed-form expressions for a short-time propagator based on the 
Milstein discretization scheme,
which has been used in Refs.~\cite{tse_estimation_2004,
pieschner_bayesian_2020}.
Drozdov \cite{drozdov_exponential_1993,
drozdov_accurate_1993,
drozdov_high-accuracy_1997,
drozdov_high-accuracy_1997-1} 
discusses several 
approximation schemes for the logarithm of the short-time propagator,
which are based on the cumulant generation function \cite{drozdov_high-accuracy_1997,
drozdov_high-accuracy_1997-1},
or a direct power-series ansatz (in powers of the short time increment $\Delta t$) in the
 exponent of the 
short-time propagator \cite{drozdov_exponential_1993,
drozdov_accurate_1993}.
Using an expansion in Hermite polynomials, A\"{i}t-Sahalia \cite{ait-sahalia_maximum_2002,
ait-sahalia_closed-form_2008}
also derives an approximation scheme for the logarithm of the short-time propagator.
One derivation of his approximation scheme involves a
 nonlinear coordinate transformation \cite{ait-sahalia_maximum_2002,
ait-sahalia_closed-form_2008},
the other \cite{ait-sahalia_closed-form_2008} uses an expansion both the
spatial increment $\Delta x$ and the time increment $\Delta t$ of the diffusion process.
 At the time of submission of the present paper, 
we became aware of the recent work of Sorkin, Ariel, and Markovich, 
who derived another approximation scheme for the 
short-time propagator \cite{sorkin_consistent_2024}; 
their scheme employs stochastic Taylor expansions \cite{kloeden_numerical_1992}
and, like A\"{i}t-Sahalia \cite{ait-sahalia_closed-form_2008}, they express their result 
in terms of Hermite polynomials.
The derivations of Drozdov and A\"{i}t-Sahalia both 
 lead to an approximate propagator that ensures the non-negativity
of the probability density exactly, but in general does not 
lead to a properly normalized probability density \cite{ait-sahalia_closed-form_2008}.
Furthermore, 
since the approximate propagators are generally in the form of an exponential
of a function (with the exception of Ref.~\cite{sorkin_consistent_2024}),
it is not straightforward to analytically evaluate 
 expectation value integrals with them.

For one-dimensional reaction coordinates, 
we here 
present a perturbation theory approach to calculate the short-time propagator
to in principle arbitrary accuracy.
The basic underlying ideas are similar to Refs.~\cite{ait-sahalia_maximum_2002,
ait-sahalia_closed-form_2008},
but our execution differs in several aspects.
First, we employ neither a nonlinear coordinate transformation, nor do we 
simultaneously expand in two variables.
Rather, our derivation uses a single linear coordinate transformation, which explicitly
 encodes the 
relative scale of the typical position increment $\Delta x$ during a short time increment $\Delta t$;
this results in a theory with a single perturbation parameter,
which is proportional to $\sqrt{\Delta t}$ (as opposed to Ref.~\cite{drozdov_exponential_1993,
drozdov_accurate_1993}, where a 
perturbation series in $\Delta t$ is used).
Second, our approach directly leads to an approximate propagator 
that is properly normalized at all orders of perturbation theory, but
can take negative values in regimes where the perturbation theory seizes to be valid.
The non-Gaussian propagators in Ref.~\cite{drozdov_exponential_1993,
drozdov_accurate_1993,
drozdov_high-accuracy_1997,
drozdov_high-accuracy_1997-1,
ait-sahalia_maximum_2002,
ait-sahalia_closed-form_2008}, on the other hand, preserve positivity exactly,
but in general are not properly normalized.
One advantage of the normalization-preserving propagator (NPP)
over the positivity-preserving propagator (PPP) is that the former
straightforwardly allows for perturbative evaluation of expectation values.
We demonstrate this by deriving explicit perturbative formulas for the 
 moments $\langle \Delta x^{n} \rangle$,
 the finite-time Kramers-Moyal coefficients, 
 and the medium entropy production rate.

We supply an accompanying python module called PySTFP \cite{PySTFP}, where 
STFP stands for \underline{s}hort-\underline{t}ime \underline{F}okker-\underline{P}lanck.
The module 
contains
 readily useable symbolic \cite{sympy} expressions for the NPP
 with a spatial integrated pointwise error that scales as $\Delta t^{9/2}$.
 %(as measured
 % by the spatially integrated absolute pointwise difference between
  % approximate  and true propagator, 
Our module furthermore includes symbolic expressions for the 
moments $\langle \Delta x^n \rangle$ for $n = 0, 1, 2, 3, 4$ up to 
an error
of the order $\Delta t^{5}$,
 the medium entropy production rate and the Gibbs entropy 
 up to an error
 of the order $\Delta t^5$,
 and the total entropy production rate up to an error of the order $\Delta t^4$.
Additionally, we provide code to reproduce all figures of this paper,
and to symbolically evaluate
all the power series we consider here (i.e.~short-time propagator,
moments, Gibbs entropy, medium entropy production rate,
total entropy production rate)
 to in principle arbitrary desired order.

The remainder of this paper is organized as follows.
In Sect.~\ref{sec:model} we first introduce the SDE and FPE that describe
one-dimensional diffusive stochastic dynamics.
We then recall a standard derivation of the Gaussian propagator in
 Sect.~\ref{sec:gaussian_propagator_derivation}.
In the subsequent Sect.~\ref{eq:perturbative_solution_of_FP},
we derive our perturbation scheme, and formulate the result 
both in a normalization-preserving and positivity-preserving representation.
In Sect.~\ref{sec:FTKM} we use the normalization-preserving representation
of the propagator to
 derive perturbative expressions for the moments
of the position increment, for the first two finite-time Kramers-Moyal coefficients,
and for the medium entropy production rate.
In Sect.~\ref{sec:numerical_example}, we illustrate all our results with a numerical example,
and in Sect.~\ref{sec:conclusions} we conclude
 by summarizing our results and discussing
their further implications.

The appendices contain more details for our derivations,
as well as additional analyses.
In particular, in App.~\ref{sec:physical_units} we give the propagator,
to second order in the perturbation theory, in both normalization- and positivity-preserving repesentations,
as well as in a midpoint-evaluation scheme.
For the latter we discuss the ratio of forward-backward path probabilities, as
related to pathwise entropy production \cite{maes_time-reversal_2003,
seifert_entropy_2005,
seifert_stochastic_2012,
cates_stochastic_2022}.
Furthermore, in App.~\ref{app:entropy} we derive power series
expansions of the Gibbs entropy, as well as the medium-
and the total entropy production rate.

\section{Diffusive stochastic dynamics}
\label{sec:model}

We consider the one-dimensional Fokker-Planck equation (FPE) \cite{risken_fokker-planck_1996,
van_kampen_stochastic_2007,
gardiner_stochastic_2009}
\begin{align}
\label{eq:fokker_planck}
\partial_t P &= - \partial_x (a P) + \partial_x^2 (D P),
\end{align}
where $D \equiv D(x)$ is a diffusivity profile, 
$a \equiv a(x)$ is a drift profile, 
and
where $P \equiv P(x,t\mid x_0,t_0)$ is the propagator, or transition (probability) density, 
to find a particle that 
starts at $(x_0,t_0)$ at the point $(x,t)$, where $t > t_0$.
We assume both drift and diffusivity are independent of time,
and seek to calculate a short-time solution to Eq.~\eqref{eq:fokker_planck}
subject to the boundary conditions
\begin{align}
\label{eq:FP_BC}
P(x,  t\mid x_0, t_0) \rightarrow 0 \qquad |x| \rightarrow \infty,
\end{align}
and the normalization condition
\begin{align}
\label{eq:propagator_normalization}
\int_{-\infty}^{\infty}dx P(x, t \mid x_0, t_0) &= 1.
\end{align}

Throughout this paper, we switch between $(x,t, x_0, t_0)$ and  
 $(\Delta x, \Delta t, x_0, t_0)$ as 
independent variables as is convenient,
where $\Delta x \equiv x-x_0$, $\Delta t \equiv t - t_0$. 
We assume $x_0$, $t_0$ as fixed and given and, unless important for the context,
 suppress explicit dependences on 
those two parameters
in the notation.

The FP Eq.~\eqref{eq:fokker_planck} is equivalent to the It\^{o}-Langevin equation
\cite{risken_fokker-planck_1996,
van_kampen_stochastic_2007,
gardiner_stochastic_2009}
\begin{align}
\label{eq:LangevinEq}
dX_t &= a(X_t)\,dt + \sqrt{2D(X_t)} \,dB_t.
\end{align}
Here, $dX_t$ is the increment of  the diffusive stochastic process $X_t$
during a time increment $dt$,
and $dB_t$ is the increment of the Wiener process.
In the It\^{o}-Langevin formulation, the delta-peak initial condition is given as
$X_{t_0} = x_0$.

\section{Gaussian short-time propagator}
\label{sec:gaussian_propagator_derivation}

For completeness and future reference, 
we now recall a standard derivation of the approximate propagator $P(x,t\mid x_0 ,t_0)$
for a short time increment $\Delta t \equiv t -t_0$ \cite{onsager_fluctuations_1953,
hunt_path_1981,
arnold_langevin_2000,
arnold_symmetric_2000,
tang_summing_2014,
cugliandolo_building_2019,
cates_stochastic_2022,
de_pirey_path_2023}.

From a stochastic Taylor expansion of Eq.~\eqref{eq:LangevinEq} it follows 
 that
 for a short time interval $\Delta t$,
we have \cite{kloeden_numerical_1992}
\begin{align}
\label{eq:short_time_langevin}
\Delta X(\Delta W) &= a(x_0) \Delta t + \sqrt{ 2D (x_0) \Delta t} \Delta W 
+ \mathcal{O}(\Delta t^\beta),
\end{align}
where $\Delta X = X_t - x_0$, and where $\Delta W$ is a
 unit normal random variable with density
\begin{equation}
P(\Delta w) = \frac{1}{\sqrt{ 2 \pi }}e^{- \Delta w^2/2}.
\end{equation}
By the notation $\Delta X(\Delta W)$ on the left-hand side of Eq.~\eqref{eq:short_time_langevin}
we emphasize that the random variable $\Delta X$ is a function of the noise increment $\Delta W$.
The error in the short-time discretization Eq.~\eqref{eq:short_time_langevin}
is of order $\Delta t^{\beta}$, 
with $\beta = 3/2$ for additive noise and $\beta = 1$ for multiplicative noise \cite{kloeden_numerical_1992};
this is the statement that the Euler-Maruyama discretization scheme
 has strong order of convergence
$\Delta t$ for additive noise, and strong order of convergence
 $\sqrt{ \Delta t}$ for multiplicative noise \cite{kloeden_numerical_1992}.

The probability density to observe an increment $\Delta x \equiv x - x_0$ follows
by a change of measure \cite{van_kampen_stochastic_2007}
\begin{equation}
\label{eq:change_of_measure}
P(\Delta x)  = \int_{-\infty}^{\infty}d\Delta w\,\delta (\Delta X(\Delta w) - \Delta x) P(\Delta w),
\end{equation}
where $\delta$ is the Dirac-delta distribution
and $\Delta X(\Delta w)$
is defined 
in Eq.~\eqref{eq:short_time_langevin}.
To evaluate the integral on the right-hand side of Eq.~\eqref{eq:change_of_measure},
we solve Eq.~\eqref{eq:short_time_langevin} for $\Delta W$,
\begin{equation}
\label{eq:short_time_langevin2}
 \Delta W(\Delta X) =\frac{ \Delta X - a(x_0) \Delta t}{\sqrt{ 2D (x_0) \Delta t}}  + \mathcal{O}(\Delta t^{\beta - 1/2}).
\end{equation}
Note that because solving Eq.~\eqref{eq:short_time_langevin} for $\Delta W$ entails dividing the
 equation by $\sqrt{\Delta t}$,
the term $\mathcal{O}(\Delta t^{\beta})$ from Eq.~\eqref{eq:short_time_langevin}
leads to a term of order $\mathcal{O}(\Delta t^{\beta-1/2})$
in Eq.~\eqref{eq:short_time_langevin2}.
From Eq.~\eqref{eq:short_time_langevin2} we obtain in particular the noise
increment $\Delta w = \Delta W(\Delta x)$ which corresponds to $\Delta x$.
Using the usual chain rule for the Dirac-delta distribution \cite{hassani_mathematical_2009},
\begin{equation}
\delta (\Delta X(\Delta w) - \Delta x) = \frac{ \delta (\Delta w - \Delta W(\Delta x))}{\left.\dfrac{d\Delta X}{d\Delta W}\right|_{\Delta W(\Delta x)}},
\end{equation}
with $\left. d\Delta X/d\Delta W \right|_{\Delta W(\Delta x)} = \sqrt{ 2 D(x_0)\Delta t}$,
the integral on the right-hand side of Eq.~\eqref{eq:change_of_measure}
then evaluates to
\begin{align}
P(\Delta x)  &= \frac{1}{\sqrt{ 2D(x_0) \Delta t}}
P(\Delta W(\Delta x))
\\
&=\frac{1}{\sqrt{ 4 \pi D(x_0) \Delta t}}
\label{eq:langevin_propagator_final}
\\
&\times
\exp\left[
 - \frac{\Delta t}{4D(x_0)}\left( \frac{\Delta x}{\Delta t} - a(x_0)\right)^2
  + \mathcal{O}(\Delta t^{\beta-1/2})
   \right]
  \nonumber,
\end{align}
where we assume that $\Delta x = \mathcal{O}(\sqrt{\Delta t})$. This is true
both if $\Delta x$ is a typical realization of the It\^{o}-Langevin Eq.~\eqref{eq:LangevinEq}
or if $\Delta x$ is a differentiable path, in which case the stronger statement
$\Delta x = \mathcal{O}(\Delta t)$ holds.

We stress that
even for additive noise,
 the exponent in the 
 short-time propagator Eq.~\eqref{eq:langevin_propagator_final}
 is only accurate to including order $\sqrt{\Delta t}$.
This is ultimately because
in solving Eq.~\eqref{eq:short_time_langevin} for $\Delta W$, 
we divide the equation by $\sqrt{\Delta t}$.

In principle the above derivation can be extended to arbitrary order by replacing
Eq.~\eqref{eq:short_time_langevin} 
with a higher-order
stochastic Taylor expansion \cite{kloeden_numerical_1992}.
This was done for additive noise in Ref.~\cite{gladrow_experimental_2021},
where the above derivation is carried out 
with the order-$\Delta t^{3/2}$ term included
 in Eq.~\eqref{eq:short_time_langevin} \cite{kloeden_numerical_1992}.
The resulting exponent in the short-time propagator from
 Ref.~\cite{gladrow_experimental_2021}
indeed differs 
from Eq.~\eqref{eq:langevin_propagator_final}
by a term of order $\Delta t$.
However, in general it does not seem practical to derive the short-time propagator
to higher accuracy via time-discretizations of the It\^{o}-Langevin Eq.~\eqref{eq:LangevinEq} directly.
The reason for this is that, 
as is apparent in Ref.~\cite{gladrow_experimental_2021}, 
higher-order stochastic Taylor expansions of the SDE
 lead to the appearance of an increasing number of correlated
random variables on the right-hand side of Eq.~\eqref{eq:short_time_langevin}.
The corresponding change of variables Eq.~\eqref{eq:change_of_measure}
then quickly becomes infeasible.
While the approach in Ref.~\cite{sorkin_consistent_2024} is based on 
stochastic Taylor expansions,
it does not follow the strategy from this section, but instead uses 
 characteristic functions.

In the next section, we consider
a derivation of the short-time propagator 
that is not based on the It\^{o}-Langevin SDE, but on the equivalent
description of the stochastic process via the FP Eq.~\eqref{eq:fokker_planck}.

\section{Perturbative short-time propagator}
\label{eq:perturbative_solution_of_FP}

\subsection{Normalization-preserving propagator}
\label{sect:NPP_derivation}
For given $x_0$, $t_0$, 
we now derive a short-time approximation for the propagator $P(x,t\mid x_0,t_0)$,
valid for $\Delta t \equiv t - t_0$ sufficiently small, and
 subject to the delta-peak initial condition
$P(x,t_0 \mid x_0,t_0) = \delta (x -x_0)$.

Intuitively, we expect the propagator to behave as follows.
For very short time $\Delta t$, the stochastic dynamics described by Eq.~\eqref{eq:fokker_planck}
is dominated by the diffusivity (as opposed to the drift).
 This is most clearly seen by considering the 
time-discretized It\^{o}-Langevin Eq.~\eqref{eq:short_time_langevin}.
For a short time increment, the random
noise term scales as $\sqrt{\Delta t}$, whereas the term that contains the deterministic drift 
scales as $\Delta t$.
For short enough time increment, the short-time propagator should thus 
be well-approximated by  the free-diffusion propagator with position-independent 
diffusivity $D(x_0)$.
Based on this intuition, 
our approach is to perturb around this free-diffusion solution, 
and to calculate corrections to it in powers
of $\sqrt{\Delta t}$.

For this we first rewrite Eq.~\eqref{eq:fokker_planck} in dimensionless form.
We fix a length scale $L$,
and note that the given initial point $x_0$ leads to a diffusivity scale $D(x_0)$.
This in turn defines 
 an associated time scale $\tD(x_0) = L^2/D(x_0)$.
The diffusivity $D(x_0)$ furthermore gives rise to a time-dependent length scale, 
namely the typical distance $R(\Delta  t)$ a freely diffusing
 particle  subject to diffusivity $D(x_0)$ travels during a short time increment $\Delta t$,
 \begin{align}
 R(\Delta t) &\equiv \sqrt{ 2 D(x_0)  \Delta t }.
 \end{align}
Using the time scale $\tD$ and length scale $R(\Delta t)$, 
we define a dimensionless time-dependent coordinate system
\begin{align}
\label{eq:tDL_def}
\tDL(\Delta t) &\equiv \frac{t-t_0}{\tD} \equiv \frac{\Delta t}{\tD},
\\ 
\label{eq:xDL_def} 
\xDL(\Delta x, \Delta t) &\equiv \frac{ x - x_0}{R(\Delta t)} \equiv \frac{ \Delta x}{R(\Delta t)},
\end{align}
where, as before, in our notation we suppress the dependence of $\tD$, $R$ on the
fixed initial position $x_0$.

With respect to the coordinates Eqs.~\eqref{eq:tDL_def}, \eqref{eq:xDL_def}, 
we rewrite the FP Eq.~\eqref{eq:fokker_planck} in 
dimensionless form as
\begin{align}
\label{eq:fokker_planck_dimensionless}
\epsDL^2 \partial_{\tDL} \PDL &= - \partial_{\xDL}\left[ \left( \epsDL \aDL - \xDL \right) \PDL \right]
+ \partial_{\xDL}^2 \left( \DDL \PDL \right),
\end{align}
with 
\begin{align}
\label{eq:def_epsDL}
\epsDL (\tDL) &\equiv \frac{R(\Delta t)}{L},\\
\label{eq:def_PDL}
\PDL(\xDL,\tDL) &\equiv R(\Delta t) P(x,t \mid x_0, t_0),\\
\label{eq:def_aDL}
\aDL(\xDL) &\equiv \frac{\tD}{L} a(x),\\
\DDL(\xDL) &\equiv \frac{D(x)}{D(x_0)},
\label{eq:def_DDL}
\end{align}
where $(\xDL,\tDL)$ are related to $(\Delta x, \Delta t)$ via Eqs.~\eqref{eq:tDL_def}, \eqref{eq:xDL_def}.

The boundary conditions Eq.~\eqref{eq:FP_BC}
become
\begin{align}
\label{eq:FP_BC_DL}
\PDL(\xDL,\tDL) \rightarrow 0 \qquad |\xDL| \rightarrow\infty,
\end{align}
and the normalization condition Eq.~\eqref{eq:propagator_normalization} is given in
 dimensionless form as
\begin{align}
\label{eq:normalization_dimensionless}
\int_{-\infty}^{\infty}d\xDL \PDL(\xDL,\tDL) &= 1.
\end{align}

By definition of $\epsDL$ it holds that $\epsDL = \sqrt{2 \tDL} \sim \sqrt{\Delta t}$, so that 
an expansion of $\PDL$ in powers of $\epsDL$ is a short-time
expansion.
Furthermore, in Eq.~\eqref{eq:xDL_def} %the rescaled spatial coordinate $\xDL$
the (on short times) dominant free-diffusion contribution to the spatial increment has been
incorporated explicitly via the denominator $R(\Delta t)$, 
so that for short time we expect
that the probability density is only non-negligible 
for values $|\xDL| \lesssim 1$.
We therefore seek a solution of Eq.~\eqref{eq:fokker_planck_dimensionless}
in the form of a power-series in $\epsDL$, 
assuming that $\xDL = \mathcal{O}(\epsDL^{0})$.

For this, we introduce a power series ansatz
\begin{align}
\label{eq:perturbative_solution_rewritten_0}
\PDL(\xDL,\tDL) &= 
\PDL^{(0)}(\xDL) \left[ 1 + \epsDL(\tDL) \Qindex{1}(\xDL) 
+ \epsDL^2(\tDL) \Qindex{2}(\xDL) + ...\right]
\\
&= \PDL^{(0)}(\xDL)\left[ \sum_{n=0}^{\infty} \epsDL^n(\tDL)\Qindex{n}(\xDL) \right]
\label{eq:perturbative_solution_rewritten_1},
\end{align}
where we identify $\Qindex{0}(\xDL) \equiv 1$ (which will be justified further below) and
where
\begin{align}
\label{eq:PDL_power_series_zero}
\PDL^{(0)}(\xDL) &= \frac{1}{\sqrt{2\pi}} \exp\left( - \frac{\xDL^2}{2}\right).
\end{align}
From the definition of $\xDL$, Eq.~\eqref{eq:xDL_def}, it is evident that 
Eq.~\eqref{eq:PDL_power_series_zero} is the free-diffusion
 solution of a FPE with spatially constant diffusivity $D(x_0)$
 and without any deterministic drift;
this is precisely the expected 
 behavior of a solution Eq.~\eqref{eq:fokker_planck}
 at asymptotically short times.
 %
 % Add reference to stochastic action for tubes paper
 
In contrast to our perturbation ansatz Eq.~\eqref{eq:perturbative_solution_rewritten_0},
Refs.~\cite{drozdov_exponential_1993,
drozdov_accurate_1993,
drozdov_high-accuracy_1997,
drozdov_high-accuracy_1997-1,
ait-sahalia_maximum_2002,
ait-sahalia_closed-form_2008} 
consider a power series ansatz in the exponent of the short-time propagator.
Furthermore, in Ref.~\cite{ait-sahalia_maximum_2002} a change of coordinates
similar to Eq.~\eqref{eq:xDL_def}
 is used, together with a nonlinear
 change in coordinates that transforms the multiplicative-noise
Eq.~\eqref{eq:xDL_def} into an additive-noise
 system. Our derivation, however,
involves no nonlinear coordinate transformations.
Conceptually our approach here is similar to 
Refs.~\cite{kappler_stochastic_2020,
kappler_sojourn_2020}
where, for an absorbing-boundary FPE,
after a linear coordinate transformation 
a systematic perturbation theory around an asymptotic
solution was developed.

 We seek to determine the functions $\Qindex{k}(\xDL)$
 such that Eq.~\eqref{eq:perturbative_solution_rewritten_0} 
 solves the dimensionless FP Eq.~\eqref{eq:fokker_planck_dimensionless},
and fulfills the  boundary conditions Eq.~\eqref{eq:FP_BC_DL}
at each order in $\epsDL$, which at order $\epsDL^k$ read
\begin{align}
\label{eq:PDL_power_series_boundary_condition}
\Qindex{k}(\xDL) \PDL^{(0)}(\xDL) &\rightarrow 0 \quad \mathrm{as}\quad|\xDL| \rightarrow \infty.
\end{align}

To derive a hierarchy of equations from Eq.~\eqref{eq:fokker_planck_dimensionless},
we first Taylor expand $a(x)$, $D(x)$ around $x_0$. 
In dimensionless units this yields
\begin{align}
\label{eq:aDL_power_series}
\aDL(\xDL,\tDL) &= \sum_{n=0}^{\infty} \An \epsDL(\tDL)^n \xDL^n,
\\
\label{eq:DDL_power_series}
\DDL(\xDL,\tDL) &= \sum_{n=0}^{\infty} \Dn \epsDL(\tDL)^n \xDL^n,
\end{align}
with
\begin{align}
\label{eq:An_def}
\An &= \tD \frac{L^{n-1}}{n!}  a^{(n)}(x_0) ,
\\
\label{eq:Dn_def}
\Dn &= \frac{L^{n}}{n!} \frac{ D^{(n)}(x_0) }{D(x_0)},
\end{align}
where a superscript $(n)$ denotes the $n$-th derivative with respect to $x$,
i.e.~$a^{(n)} \equiv \partial_x^n a$, $D^{(n)} \equiv \partial_x^n D$.
We note that from Eq.~\eqref{eq:Dn_def} we have $\Dzero \equiv 1$.

We substitute 
the power series expansions Eqs.~\eqref{eq:perturbative_solution_rewritten_1}, 
\eqref{eq:aDL_power_series}, \eqref{eq:DDL_power_series},
 into the dimensionless FP Eq.~\eqref{eq:fokker_planck_dimensionless},
and demand that the resulting equation hold at each power of $\epsDL$ separately.
This yields a hierarchy of equations, which at order $\epsDL^k$ is given by
\begin{align}
\label{eq:dimensionless_hierarchy_Q}
\partial_{\xDL}^2 &\Qindex{k}
- \xDL \partial_{\xDL} \Qindex{k}
- k \Qindex{k}
\\ 
&= 
\sum_{l=0}^{k-1}
\Aindex{l}
\tilde{\mathcal{L}}_{\mathcal{A},l}
\Qindex{k-1-l}
- 
\sum_{l=1}^{k}
\Dindex{l}
\tilde{\mathcal{L}}_{\mathcal{D},l}
\Qindex{k-l}.
\nonumber
\end{align}
Here, we define the sums on the right-hand side as zero if the upper summation 
bound is smaller than the lower summation bound,
and
 the linear differential operators $\tilde{\mathcal{L}}_{\mathcal{A},l}$,
$\tilde{\mathcal{L}}_{\mathcal{D},l}$ are given by
\begin{align}
\tilde{\mathcal{L}}_{\mathcal{A},l} 
\Qindex{k-1-l}
&= 
\partial_{\xDL}\left( \xDL^{l} \Qindex{k-1-l}\right) - \xDL^{l+1} \Qindex{k-1-l},
\label{eq:operator_L_A}
\\
\label{eq:operator_L_D}
\tilde{\mathcal{L}}_{\mathcal{D},l} \Qindex{k-l}
&= 
\partial_{\xDL}^2\left( \xDL^{l} \Qindex{k-l}\right)
 - 2\xDL \partial_{\xDL}\left( \xDL^{l} \Qindex{k-l} \right)
 \\ & \qquad\nonumber
 + \xDL^l \left( \xDL^2 -1 \right) \Qindex{k-l}.
\end{align}
Since on the right-hand side of  Eq.~\eqref{eq:dimensionless_hierarchy_Q}
only the $\Qindex{l}$ with $l < k$ appear, we can solve the equation for $\Qindex{k}$ 
recursively for ascending $k$.
In App.~\eqref{app:solution_recursive} we provide a recursive solution approach to
 Eq.~\eqref{eq:dimensionless_hierarchy_Q},
and prove that $\Qindex{k}$ is a polynomial in $\xDL$ of order at most $3k$.
Since $\Qindex{k}$ is a polynomial and $\PDL^{(0)}$ is a Gaussian,
the boundary conditions Eq.~\eqref{eq:PDL_power_series_boundary_condition}
are fulfilled for every $k$.

The lowest order expressions for $\Qindex{k}$ we derive are
\begin{align}
\label{eq:Qindex0}
\Qindex{0} &= 1,
\\
\label{eq:Qindex1}
\Qindex{1} &= \frac{\xDL}{4}\left( 2 \Aindex{0} - 3 \Dindex{1} + \Dindex{1}\xDL^2\right),
\\
\label{eq:Qindex2}
\Qindex{2} &=
\left( \frac{\Aindex{0}^2}{8}+ \frac{\Aindex{1}}{4} \right)\left( \xDL^2 - 1\right)
+ 
\frac{ \Aindex{0} \Dindex{1} }{8}  \left( \xDL^4 - 5\xDL^2 +2 \right)
\\ & \quad
+
\frac{\Dindex{1}^2}{32} \left( \xDL^6 - 11 \xDL^4 + 21 \xDL^2 - 3\right)
+
\frac{\Dindex{2}}{12} \left(  2 \xDL^4 -9 \xDL^2 + 3\right).
\nonumber
\end{align}
In our python module PySTFP \cite{PySTFP} we include the symbolic
expressions for $\Qindex{0}$, $\Qindex{1}$, ..., $\Qindex{8}$,
as well as code to  solve Eq.~\eqref{eq:dimensionless_hierarchy_Q} recursively
to arbitrary desired order.

To use the perturbative solution Eq.~\eqref{eq:perturbative_solution_rewritten_0}
in practice, we truncate the infinite power series at a finite number of terms $K \in \mathbb{N}_0$, 
to get
\begin{align}
\label{eq:perturbative_solution_K}
\PDL_K(\xDL,\tDL) &= 
 P^{(0)}(\xDL)\left[ \sum_{n=0}^{K} \epsDL^n(\tDL)\Qindex{n}(\xDL) \right]
 + \mathcal{O}(\epsDL^{K+1}),
\end{align}
where from the definition of $\epsDL$ we have
 $\epsDL^{K+1} \sim \Delta t^{(K+1)/2}$.
 
 In App.~\ref{app:derivation} we show that 
in the form Eq.~\eqref{eq:perturbative_solution_rewritten_0},
our perturbative solution of the FPE conserves probability exactly, i.e.~that
\begin{align}
\label{eq:global_prob_cons}
\int_{-\infty}^{\infty}d\xDL \PDL_K(\xDL,\tDL) &= 1
\end{align}
for any $K$.
We therefore refer to Eq.~\eqref{eq:perturbative_solution_K}
as the normalization-preserving propagator (NPP).

\subsection{Positivity-preserving propagator}
\label{sec:PPP}

While 
the NPP Eq.~\eqref{eq:perturbative_solution_K}
conserves probability exactly,
for large enough $\epsDL$, $\xDL$
it 
can violate the non-negativity condition
\begin{equation}
\PDL(\xDL,\tDL) \geq 0,
\label{eq:non_negativity}
\end{equation}
 which any continuous probability density needs
to fulfill.

We can rewrite Eq.~\eqref{eq:perturbative_solution_K} in a form
 that is manifestly positive by expressing the 
sum of the $\epsDL^{k}\Qindex{k}$
in  Eq.~\eqref{eq:perturbative_solution_K} in 
exponential form, as
\begin{align}
\label{eq:perturbative_solution_rewritten_3}
\PDL_K(\xDL,\tDL) &= \frac{1}{\sqrt{2 \pi}}
\exp\left\{ - \frac{ \xDL^2}{2}
\right. \\ & \qquad \left. \nonumber
+ \ln\left[ 1 + \sum_{n=1}^{K} 
\epsDL^n(\tDL) \Qindex{n}(\xDL) + \mathcal{O}(\epsDL^{K+1})\right]
\right\},
\end{align}
where we use Eqs.~\eqref{eq:PDL_power_series_zero}, \eqref{eq:Qindex0}.
Using the Taylor series of $\ln(1+z)$ around $z=0$, 
\begin{equation}
\label{eq:ln_taylor}
\ln(1+z) = - \sum_{m=1}^{K} \frac{ (-z)^m}{m} + \mathcal{O}(z^{K+1}),
\end{equation}
 we further rewrite the  exponent in Eq.~\eqref{eq:perturbative_solution_rewritten_3} as
\begin{align}
\PDL_K(\xDL,\tDL) &= \frac{1}{\sqrt{2 \pi}}
\exp\left\{ - \frac{ \xDL^2}{2}
- \sum_{m=1}^{K}\frac{1}{m}\left[ -\sum_{n=1}^{K} 
\epsDL^n(\tDL) \Qindex{n}(\xDL) \right]^m
\nonumber
\right. \\ & \qquad\qquad \left. \vphantom{\frac{1}{2}}
\label{eq:perturbative_solution_rewritten_4}
+\mathcal{O}(\epsDL^{K+1})
\right\}.
\end{align}
This expression can be evaluated to any desired order $\epsDL^{K}$ in the exponent.
As an exponential it is manifestly positive, so that 
we refer to Eq.~\eqref{eq:perturbative_solution_rewritten_4} as the positivity-preserving
propagator (PPP).

The NPP Eq.~\eqref{eq:perturbative_solution_K}
and the PPP
Eq.~\eqref{eq:perturbative_solution_rewritten_4} are only 
equivalent to order $\epsDL^K$
for small enough $\epsDL$, since Eq.~\eqref{eq:ln_taylor} only
converges for $|z| < 1$.
Beyond this, the two representations of the 
 perturbative solution have different properties.
Equation \eqref{eq:perturbative_solution_K} preserves probability exactly,
but can violate the condition that a probability density is always nonnegative,
Eq.~\eqref{eq:non_negativity}.
On the other hand,
 Eq.~\eqref{eq:perturbative_solution_rewritten_4} is manifestly positive,
but in general does not exactly preserve the probability normalization.

Even more, the PPP Eq.~\eqref{eq:perturbative_solution_rewritten_4}
can violate the boundary conditions Eq.~\eqref{eq:FP_BC_DL}.
This is because
the exponent in Eq.~\eqref{eq:perturbative_solution_rewritten_4}
 is a polynomial in $\xDL$, 
which for large $|\xDL|$ will be dominated by its highest power. 
Depending on the diffusivity $D$, drift $a$, and initial condition $x_0$,
the prefactor of this highest power in $\xDL$ might be such that 
the exponent in Eq.~\eqref{eq:perturbative_solution_rewritten_4} approaches positive infinity
as $\xDL \rightarrow \infty$ or $\xDL \rightarrow - \infty$,
so that $\PDL \rightarrow \infty$ in that limit.
In App.~\ref{sec:physical_units} we discuss this point further,
using explicit second-order perturbation theory results.

Both issues (violating nonnegativity and violating normalization)
only manifest themselves for values of $\Delta x$, $\Delta t$,
where our perturbation theory breaks down.
Nonetheless, depending on the context it is beneficial to choose 
either of the forms
Eq.~\eqref{eq:perturbative_solution_K}, 
\eqref{eq:perturbative_solution_rewritten_4}.
In particular, as we discuss in Sect.~\ref{sec:FTKM} below,
to %calculate the moments $\langle \Delta x^n\rangle$
calculate expectation values of observables
it is preferable to use the NPP Eq.~\eqref{eq:perturbative_solution_K}.

\subsection{Breakdown of the perturbative solution}

Our power series ansatz Eq.~\eqref{eq:perturbative_solution_rewritten_0} is based on 
 perturbing around a free-diffusion solution.
For very large drift or quickly varying diffusivity profile we therefore expect that a
large number of terms are needed to achieve an given accuracy for the 
perturbative short-time propagator.

More generally, we expect the perturbation ansatz Eq.~\eqref{eq:perturbative_solution_rewritten_0} to break
 down once the perturbation parameter
$\epsDL$ is not small anymore, i.e.~once $\epsDL \gtrsim 1$.
We therefore estimate the breakdown time $\Delta t_b$ of our perturbative ansatz via
$ \epsDL(\tDL = \Delta t_b/\tD) = 1/2$, which according to Eq.~\eqref{eq:def_epsDL} is equivalent to
\begin{align}
\label{eq:temporal_breakdown_condition}
\frac{\Delta t_b}{\tD} &= \frac{1}{8}.
\end{align}

\subsection{$L^1$-error for propagator}

We now quantify the quality of a
normalization-preserving approximation
of the exact propagator.
We only consider the NPP here because it
readily allows
us to evaluate expectation value integrals perturbatively, since the integrand 
is a product of a Gaussian and a polynomial.
By contrast, the PPP leads to non-Gaussian exponentials
which, as we discussed in Sect.~\ref{sec:PPP},  might not even be normalizable.
For our numerical example in Sect.~\ref{sec:numerical_example} below, 
we in App.~\ref{sec:physical_units} also discuss 
the errors for the PPP, 
both in the form Eq.~\eqref{eq:perturbative_solution_rewritten_4}
and for
 a midpoint-discretization scheme.

To measure how well an exact probability
density $P^{\mathrm{e}}$ is approximated by an estimate probability density $P$,
we consider the $L^1$-error $E(\Delta t \mid x_0)$.
This error is defined as
\begin{align}
\label{eq:L1_error}
E(\Delta t \mid x_0) &\equiv || P - P^{\mathrm{e}}||_1
\equiv
\int_{-\infty}^{\infty}dx\, E_p(x,\Delta t\mid x_0),
\end{align}
where the
pointwise error $E_p$ of the continuous probability densities we consider
 is given by
\begin{align}
\label{eq:pointwise_error}
E_p(x,\Delta t\mid x_0) &\equiv | P(x,t \mid x_0,t_0) - P^{\mathrm{e}}(x,t\mid x_0,t_0) |.
\end{align}

By direct substitution, the error estimates Eq.~\eqref{eq:langevin_propagator_final} of the GP
 and Eq.~\eqref{eq:perturbative_solution_K} of the NPP
 yield the corresponding scaling of the $L^1$ errors Eq.~\eqref{eq:L1_error}, i.e.~
\begin{align}
\label{eq:EGP}
E_{\mathrm{GP}}(\Delta t \mid x_0) &\sim \Delta t^{\beta - 1/2},
\\
\label{eq:ENPP}
E_{\mathrm{NPP}}(\Delta t \mid x_0) &\sim \Delta t^{(K+1)/2},
\end{align}
where as before $\beta = 1$ for multiplicative noise and $\beta =3/2$
for additive noise, and the integer $K$ is the truncation order
in Eq.~\eqref{eq:perturbative_solution_K}.
Note that Eq.~\eqref{eq:EGP} shows that, in general,
the GP approximates the true transition density only to sublinear
order in the time increment $\Delta t$.

\section{Observables}
\label{sec:FTKM}

\subsection{Moments}

The NPP Eq.~\eqref{eq:perturbative_solution_rewritten_1}
allows us to evaluate any expectation value perturbatively. 
For example, for $n \in \mathbb{N}_0$
we obtain a perturbation series for the $n$-th moment  as
\begin{align}
\left\langle \Delta x^n \right\rangle%\right|_{X_0 = x_0}
&= R^n \left\langle \xDL^n\right\rangle
= R^n\sum_{k=0}^{\infty} \epsDL^k \left\langle \xDL^n\right\rangle^{(k)}
\\ &=
 R^n\sum_{\substack{k=0\\k+n\mathrm{~even}}}^{\infty} \epsDL^k \left\langle \xDL^n\right\rangle^{(k)},
 \label{eq:n_th_moment_2}
\end{align}
where
\begin{align}
\label{eq:xDL_n_k}
\langle \xDL^n \rangle^{(k)} &\equiv \int_{-\infty}^{\infty}d\xDL\,\xDL^n \Qindex{k}(\xDL) \PDL^{(0)}(\xDL),
\end{align}
and where we use that $\langle \xDL^{n}\rangle^{(k)}  = 0$ whenever $n+k$ is odd,
as shown in App.~\ref{app:parity}.
The form of Eq.~\eqref{eq:xDL_n_k} demonstrates why for the moments it is preferable to use the
NPP over the PPP. 
Namely, the integral in Eq.~\eqref{eq:xDL_n_k} is a product of a polynomial in $\xDL$ and 
 a Gaussian, 
which can readily be evaluated analytically via repeated integration by parts.
On the other hand, 
the evaluation of expectation value integrals
using the PPP Eq.~\eqref{eq:perturbative_solution_rewritten_4} is not as straightforward,
for example since the truncated PPP need not even be normalizable.
 %in precisely our NPP form).

Since $R \sim \epsDL  \sim \Delta t^{1/2}$, it follows from 
Eq.~\eqref{eq:n_th_moment_2} that
\begin{align}
\label{eq:moments_order}
\langle \Delta x^n \rangle &= \begin{cases}
\mathcal{O}(\Delta t^{n/2})& n\mathrm{~even},\\
\mathcal{O}(\Delta t^{(n+1)/2})& n\mathrm{~odd},
\end{cases}
\end{align}
so that to linear order
in $\Delta t$,
the only nonvanishing moments are $\langle \Delta x^0 \rangle \equiv 1$, $\langle \Delta x\rangle$, $\langle \Delta x^2\rangle$;
this is of course required by the Pawula Theorem \cite{pawula_approximation_1967,
van_kampen_stochastic_2007}.

Using our NPP to order $K$,
we can evaluate the $n$-th moment perturbatively.
More explicitly, substituting  Eq.~\eqref{eq:perturbative_solution_K}
into the series Eq.~\eqref{eq:n_th_moment_2},
we obtain 
\begin{align}
\left\langle \Delta x^n \right\rangle%\right|_{X_0 = x_0}
&= 
 R^n \left[\,\sum_{\substack{k=0\\k+n\mathrm{~even}}}^{K} 
 \epsDL^k \left\langle \xDL^n\right\rangle^{(k)} + \mathcal{O}(\epsDL^{\,\gamma})
 \right].
\end{align}
where $\gamma = K+1$ if $n+K$ is odd, and $\gamma = K+2$ if $n+K$ is even.
Recalling that $R \sim \epsDL \sim \Delta t^{1/2}$, 
we thus have that
\begin{align}
\label{eq:n_th_moment_truncated_error}
\left\langle \Delta x^n \right\rangle%\right|_{X_0 = x_0}
&= 
 R^n \sum_{\substack{k=0\\k+n\mathrm{~even}}}^{K} 
 \epsDL^k \left\langle \xDL^n\right\rangle^{(k)} 
 \\ & \qquad 
 +
 \begin{cases} 
 	\mathcal{O}(\Delta t^{(n+K+2)/2}) &  \mathrm{if}~n+K~\mathrm{\textcolor{black}{even}},\\
 	\mathcal{O}(\Delta t ^{(n+K+1)/2}) &  \mathrm{if}~n+K~\mathrm{\textcolor{black}{odd}}.
\end{cases}
\nonumber
\end{align}

Using Eq.~\eqref{eq:xDL_n_k}
we can evaluate each moment to any desired order,
once the necessary 
polynomials $\Qindex{k}(\xDL)$ have been derived using the recursive scheme from Sect.~\ref{sect:NPP_derivation};
in App.~\ref{app:moments}, we print the resulting perturbation expansions of $\langle \Delta x^n \rangle$ to 
order $\Delta t^3$ for $n = 0, 1, 2, 3$.
\textcolor{black}{In the module PySTFP we provide analytical expressions for the moments 
$\langle \Delta x^n \rangle$ for
$n = 0, 1, ..., 4$ up to errors of the order $\Delta t^{5}$ \cite{PySTFP}.}

\subsection{Finite-time Kramers-Moyal coefficients}

From Eq.~\eqref{eq:n_th_moment_truncated_error}, we in particular obtain
 finite-time generalizations to the Kramers-Moyal coefficients.
 To leading orders these follow  as
\begin{align}
\label{eq:alpha_one}
\alpha_1(x_0,\Delta t) &\equiv 
\frac{\langle \Delta x \rangle}{\Delta t} 
\\
 \nonumber
&= a + \frac{ \Delta t}{2} \left[ a \partial_x a + D \partial_x^2a \right] 
+ \mathcal{O}(\Delta t^2),
\\
\label{eq:alpha_two}
\alpha_2(x_0,\Delta t) &\equiv 
\frac{\langle \Delta x^2 \rangle}{\Delta t} 
\\
&= 
2D + \Delta t \left[
 a^2
 + a \partial_x D 
 + 2 (\partial_x a)D +D \partial_x^2 D
\right] 
 \nonumber
\\ &\quad+ \mathcal{O}(\Delta t^2),
 \nonumber
\end{align}
where $a$, $D$, and their derivatives are evaluated at $x_0$.
From Eqs.~\eqref{eq:alpha_one}, \eqref{eq:alpha_two} it is apparent that in the limit
$\Delta t \rightarrow 0$, the usual Kramers-Moyal coefficients  \cite{van_kampen_stochastic_2007,
 gardiner_stochastic_2009} are recovered.
According to Eqs.~\eqref{eq:alpha_one},
\eqref{eq:alpha_two} the first two moments both scale as $\Delta t$ to leading order,
which is in agreement with Eq.~\eqref{eq:moments_order}.
Our Eqs.~\eqref{eq:alpha_one}, 
\eqref{eq:alpha_two} are identical to those previously
derived in the literature \cite{friedrich_comment_2002,
gottschall_definition_2008,
rydin_gorjao_arbitrary-order_2021}.

\subsection{Medium entropy production rate}

As another observable, we now consider the ensemble 
medium entropy production \cite{seifert_entropy_2005,
seifert_stochastic_2012,
seifert_stochastic_2019} 
\begin{align}
\label{eq:def_Sm}
\dot{S}^{\mathrm{m}}(t) 
&\equiv
\int_{-\infty}^{\infty}dx\,\frac{ j(x,t)}{D(x)}\left[
a(x) - (\partial_x D)(x)
 \right],
\end{align}
with the standard FP probability flux
\begin{align}
\label{eq:def_j}
j &\equiv aP - \partial_x (D P).
\end{align}
Physically speaking, 
the ensemble medium entropy production Eq.~\eqref{eq:def_Sm}
describes the rate at which the reaction coordinate dissipates
energy into the heat bath \cite{seifert_entropy_2005} 
(which is modeled via the random force term 
in the It\^{o}-Langevin Eq.~\eqref{eq:LangevinEq}).

As we discuss in detail in App.~\ref{app:entropy},
by substituting our perturbative propagator Eq.~\eqref{eq:perturbative_solution_rewritten_1} into
Eq.~\eqref{eq:def_Sm} we obtain a short-time perturbation series 
for $\dot{S}^{\mathrm{m}}$. To leading
order we have
\begin{align}
\label{eq:Sm_dot_leading_order}
\dot{S}^{\mathrm{m}} 
=&
 \frac{1}{D}\left( a - \partial_x D\right)^2  + \partial_x a -  \partial_x^2 D
+ \mathcal{O}(\Delta t),
\end{align}
where $a$, $D$ and their derivatives are evaluated at $x_0$.
In our python module PySTFP \cite{PySTFP} we give the symbolic expression for
$\dot{S}^{\mathrm{m}}$ up to order $\Delta t^4$.

For systems that
 allow for a zero-flux steady state, 
 which includes (non-periodic) one-dimensional systems with
  a drift that confines the particle to a finite spatial domain,
we in App.~\ref{eq:app_medium_potential} recall that the
 physical interpretation of Eq.~\eqref{eq:def_Sm} can be made more explicit
 \cite{sekimoto_stochastic_2010,
 seifert_stochastic_2019}.
Briefly, for such systems we can define a potential $U$ such that the steady state
is a Boltzmann distribution with respect to $U$.
The medium entropy production Eq.~\eqref{eq:def_Sm} can then be rewritten in terms of $U$ as
\cite{sekimoto_stochastic_2010,seifert_stochastic_2019}
\begin{align}
\label{eq:Sm_potential}
\dot{S}^{\mathrm{m}}(t)  
=&
-\partial_t \langle U \rangle,
\end{align}
i.e.~as the negative rate of change of the mean potential (internal energy) of the system.
Since any change in potential has to happen through exchange of energy with the heat bath,
we can interpret Eq.~\eqref{eq:Sm_dot_leading_order} equivalently as describing the 
mean loss in internal energy, or the mean energy
dissipated into the heat bath \cite{seifert_entropy_2005,
sekimoto_stochastic_2010,
seifert_stochastic_2019}.

While we consider the medium entropy production rate here, 
we in App.~\ref{app:entropy} 
 provide perturbative evaluations of
 the Gibbs entropy, as well as the total entropy production rate, from the 
stochastic thermodynamics literature \cite{seifert_entropy_2005,
seifert_stochastic_2012}.

\subsection{Error for expectation values}
\label{eq:error_estimates_observables}

For a general observable $f(x,t, x_0,t_0)$
we consider the error in the instantaneous expectation value, defined by
\begin{align}
\label{eq:Error_expectation_value}
E_f (\Delta t \mid x_0) &\equiv | \langle f \rangle - \langle f \rangle^{\mathrm{e}}|,
\end{align}
where 
\begin{align}
\langle f \rangle &\equiv \int_{-\infty}^{\infty}dx f(x,t,x_0,t_0)P(x, t\mid x_0,t_0),
\\
\langle f \rangle^{\mathrm{e}} &\equiv \int_{-\infty}^{\infty}dx f(x,t,x_0,t_0)P^{\mathrm{e}}(x, t\mid x_0,t_0).
\end{align}
with $P$, $P^{\mathrm{e}}$ the approximate and exact propagator, respectively.

For the error Eq.~\eqref{eq:Error_expectation_value}
we first consider the special case where $f \equiv \langle \Delta x^n\rangle$ for some 
positive integer $n$.
From Eq.~\eqref{eq:n_th_moment_truncated_error}
we then obtain that for the NPP at order $K$ it holds that
\begin{align}
\label{eq:EPPn}
E_{\mathrm{NPP},{\langle \Delta x^n \rangle}}(\Delta t\mid x_0) &\sim
 \begin{cases} 
 	\Delta t^{(n+K+2)/2} &  \mathrm{if}~n+K~\mathrm{\textcolor{black}{even}},\\
	\Delta t ^{(n+K+1)/2} &  \mathrm{if}~n+K~\mathrm{\textcolor{black}{odd}}.
\end{cases}
\end{align}
Thus, for the first two finite-time Kramers-Moyal coefficients
we have
\begin{align}
\label{eq:EPP_one}
E_{\mathrm{NPP},\alpha_{1}}(\Delta t\mid x_0) &\sim
 \begin{cases} 
	\Delta t^{(K+1)/2} &  \mathrm{if}~K~\mathrm{\textcolor{black}{odd}},\\
 	\Delta t ^{K/2} &  \mathrm{if}~K~\mathrm{\textcolor{black}{even}},
\end{cases}
\\
\label{eq:EPP_two}
E_{\mathrm{NPP},\alpha_{2}}(\Delta t\mid x_0) &\sim
 \begin{cases} 
	\Delta t^{K/2 + 1} &  \mathrm{if}~K~\mathrm{\textcolor{black}{even}},\\
	\Delta t ^{(K+1)/2} &  \mathrm{if}~K~\mathrm{\textcolor{black}{odd}},
\end{cases}
\end{align}
where 
$\alpha_1$, $\alpha_2$ are defined by Eqs.~\eqref{eq:alpha_one}, \eqref{eq:alpha_two},
and where we use the perturbative expectation values 
from Eq.~\eqref{eq:n_th_moment_truncated_error}.

For a general analytical function $f$,
the expectation value $\langle f \rangle$ can be related
to the moments Eq.~\eqref{eq:n_th_moment_2} 
via a two-dimensional Taylor expansion of $f$ around 
 $(x,t) = (x_0,t_0)$.
 For example, to calculate $\langle f \rangle$ to linear order in the
 timestep, we obtain
\begin{align}
\label{eq:observable_taylor}
\langle f\rangle &= f 
+ (\partial_x f)\langle \Delta x\rangle
+ \frac{\partial_x^2 f}{2} \langle \Delta x^2 \rangle
\\ & \qquad \nonumber
+ (\partial_t f)\Delta t
+ \langle \mathcal{O}(\Delta x^3)\rangle
+ \mathcal{O}(\Delta t^2),
\end{align}
where $f$ and its derivatives on the right-hand side are all 
evaluated at $(x ,t,x_0,t_0) = (x_0,t_0,x_0,t_0)$
and where we use that the NPP conserves probability exactly.
Using Eqs.~\eqref{eq:alpha_one}, \eqref{eq:alpha_two}, \eqref{eq:moments_order}
we get an explicit expression for Eq.~\eqref{eq:observable_taylor} 
in terms of diffusivity and drift as
\begin{align}
\langle f(\Delta x,\Delta t,x_0)\rangle &= f 
+ \Delta t \left[  a \partial_x f
+  D\partial_x^2 f
+ \partial_t f\right]
\nonumber
\\ & \qquad 
+ \mathcal{O}(\Delta t^2),
\label{eq:observable_taylor2}
\end{align}
where $a$, $D$, $f$ and its derivatives are evaluated at $(x,t) =(x_0,t_0)$.

Similarly, higher-order approximations to $\langle f \rangle$ are obtained 
by higher-order Taylor expansion of $f$ and subsequent perturbative
calculation of the moments $\langle \Delta x^n \rangle$ that emerge from the
Taylor expansion.
We now discuss %in more detail 
how to evaluate the expectation value $\langle f \rangle$ up to an error of order $\Delta t^{\alpha+1}$, 
i.e.~to accuracy $\Delta t^{\alpha}$.
From Eq.~\eqref{eq:moments_order} we see that
only the moments $\langle \Delta x^0 \rangle \equiv 1$,
$\langle \Delta x^1 \rangle$, 
...,
$\langle \Delta x^{2\alpha} \rangle$
contribute to any expression at order $\Delta t^{\alpha}$;
we therefore need to Taylor expand $f$ up to order $2 \alpha$.
For a fixed order $K$ of the NPP, the errors of the moments 
$\langle \Delta x^0 \rangle \equiv 1$,
$\langle \Delta x \rangle$,
$\langle \Delta x^2 \rangle$,
...,
$\langle \Delta x^{2\alpha} \rangle$
are then given by Eq.~\eqref{eq:EPPn}.
Among the relevant moments, the worst accuracy 
is realized by the first moment, $\langle \Delta x \rangle$.
In fact, from Eq.~\eqref{eq:EPPn}
 it follows that, to obtain $\langle \Delta x \rangle$ with an
 error that scales as $\Delta t^{\alpha+1}$,
we need $K \geq \alpha$.
Summing up, to calculate $\langle f \rangle$ up to an error that scales as $\Delta t^{\alpha+1}$,
we need to consider the spatial Taylor expansion of $f$ up to order $\Delta t^{2\alpha}$,
and need to know the NPP to order at least $K = \alpha$ to evaluate the the expectation values.
Equation \eqref{eq:observable_taylor2} represents the case $\alpha = 1$, 
where we needed to consider the first and the second moment (since $2 \alpha = 2$),
and where we need to know the NPP to order at least $K = \alpha = 1$ to evaluate 
the expectation values.

To estimate the error in expectation value for the 
GP Eq.~\eqref{eq:langevin_propagator_final},
we first %consider the finite-time Kramers-Moyal coefficients.
note that
from direct calculation we have 
 %this 
 that this propagator  predicts 
 the first two moments as
$\langle \Delta x \rangle_{\mathrm{GP}} = a(x_0) \Delta t$, 
$\langle \Delta x^2 \rangle_{\mathrm{GP}} = 2D(x_0) \Delta t + (\partial_x a)^2 \Delta t^2$.
Upon comparison with Eqs.~\eqref{eq:alpha_one}, \eqref{eq:alpha_two},
we conclude that for $n=1,2$ we have
\begin{align}
\label{eq:EGPn}
E_{\mathrm{GP},{\langle \Delta x^n \rangle}}(\Delta t\mid x_0) &\sim \Delta t^2.
\end{align}
Consequently, we obtain the corresponding error
 in the first two finite-time Kramers-Moyal coefficients as
\begin{align}
\label{eq:GP_KM}
E_{\mathrm{GP},\alpha_{n}}(\Delta t\mid x_0) &\sim \Delta t.%\\
\end{align}
From Eqs.~\eqref{eq:observable_taylor}, \eqref{eq:EGPn}, we furthermore
conclude that for the GP and an analytical function $f$ it in general holds that
\begin{align}
\label{eq:error_GP_f}
E_{\mathrm{GP},f}(\Delta t\mid x_0) &\sim \Delta t^2.
\end{align}
Note that this scaling is not in contradiction with Eq.~\eqref{eq:GP_KM};
this is because $\alpha_1$, $\alpha_2$ are not analytic functions in $\Delta t$,
and thus cannot be represented in the form Eq.~\eqref{eq:observable_taylor}.

Equations \eqref{eq:EGP}, \eqref{eq:error_GP_f} demonstrate that,
while for a general It\^{o}-Langevin equation with multiplicative noise
the error in the GP Eq.~\eqref{eq:langevin_propagator_final}
scales to leading order as $\Delta t^{1/2}$,
this propagator
it is still sufficient to evaluate expectation values with order-$\Delta t$ accuracy.

\section{Numerical example}
\label{sec:numerical_example}

\subsection{System}

\begin{figure}[ht]
\centering \includegraphics[width=1\columnwidth]{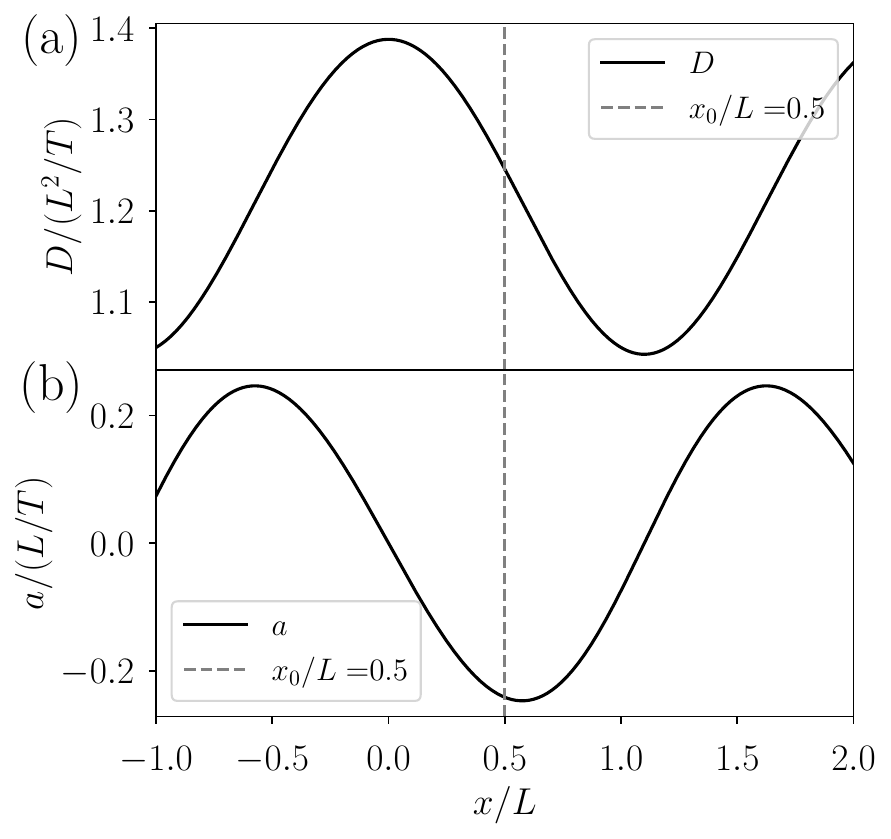} 
\caption{ \label{fig:diffusivity_and_drift} 
\textbf{Diffusivity and drift profile for the numerical example 
in Sect.~\ref{sec:numerical_example}.}
The black lines show the (a) diffusivity and (b) drift profile defined in 
App.~\ref{app:example_system} as a function of the position $x$. 
The vertical dashed line denotes the initial position of the particle for which we consider 
the short-time propagator.
}
\end{figure}

For the length scale $L$ and a time scale $T$, 
we now consider an explicit example system. 
As explained in more detail
in App.~\ref{app:example_system},
we construct diffusivity and drift profiles such that
 we can evaluate the analytical propagator
exactly.
We show the resulting
 diffusivity 
and drift profiles in Fig.~\ref{fig:diffusivity_and_drift}. 
Unless stated otherwise we in the following consider 
the initial condition $x_0/L = 0.5$, which is shown 
in Fig.~\ref{fig:diffusivity_and_drift} as vertical gray dashed line.
For the diffusivity, drift, and initial position that we consider here 
the breakdown lagtime Eq.~\eqref{eq:temporal_breakdown_condition} 
evaluates to
$\Delta t_b/T \approx 0.10$,

\subsection{Normalization-preserving propagator}

\begin{figure*}[ht]
\centering \includegraphics[width=1\textwidth]{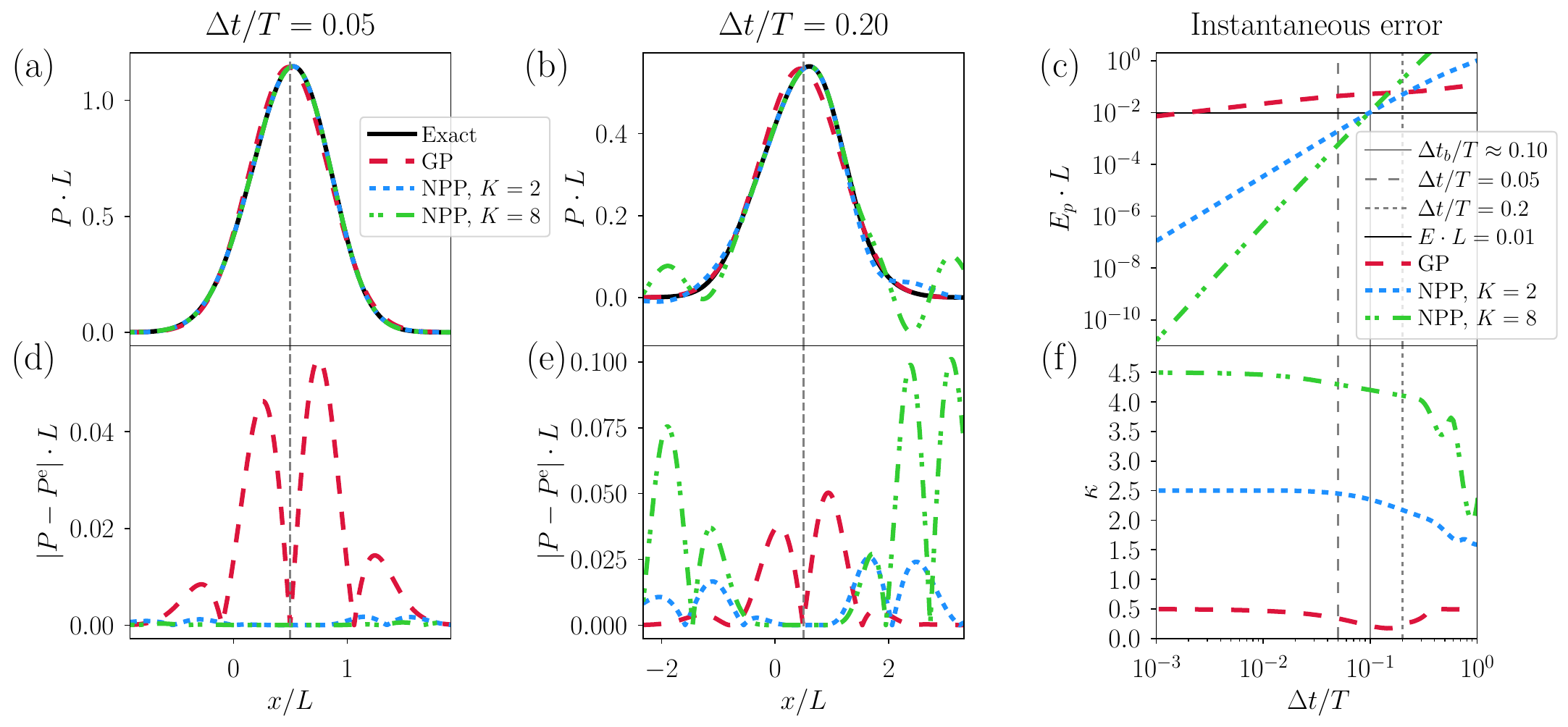} 
\caption{
 \label{fig:comparison_various_orders} 
\textbf{Comparison of exact Fokker-Planck solution with
 various approximate solutions.}
Throughout this figure, data pertaining to the Gaussian 
propagator (GP) Eq.~\eqref{eq:langevin_propagator_final}
is shown as red dashed line;
data pertaining to  the normalization-preserving propagator (NPP) Eq.~\eqref{eq:perturbative_solution_K}
is  shown for $K=2$ as blue dash-dotted lines,
and for $K=8$ as a green dotted lines.
For all data we use the the drift and diffusivity from Fig.~\ref{fig:diffusivity_and_drift},
as well as the initial condition $x_0/L = 0.5$.
In subplots (a), (b) we plot the exact solution $P^e$ to the FP Eq.~\eqref{eq:fokker_planck} (black solid line),
as well as various approximations.
In subplots (d), (e) we plot the respective pointwise error
 of the approximate propagators, as defined in Eq.~\eqref{eq:pointwise_error}.
While in subplots (a), (d) we consider propagators for the lagtime $\Delta t/T = 0.05$,
in subplots (b), (e) we consider $\Delta t/T = 0.2$.
The legend from subplot (a) is valid for subplots (a), (b), (d), (e).
In subplot (c) we show the $L^1$ error Eq.~\eqref{eq:L1_error} of the approximate 
propagators as a function of the lagtime $\Delta t/T$;
in subplot (f) we plot the corresponding local exponents Eq.~\eqref{eq:running_exponent}.
The vertical lines in subplots (c), (f) indicate the lagtimes used for subplots (a), (b), (d), (e),
as well as the breakdown time $\Delta t_b/T \approx 0.10$ defined via 
Eq.~\eqref{eq:temporal_breakdown_condition}.
The horizontal line in subplot (c) indicates the error value $E_p \cdot L = 0.01 = 1\%$.
}
\end{figure*}

\begin{figure*}[ht]
\centering \includegraphics[width=1\textwidth]{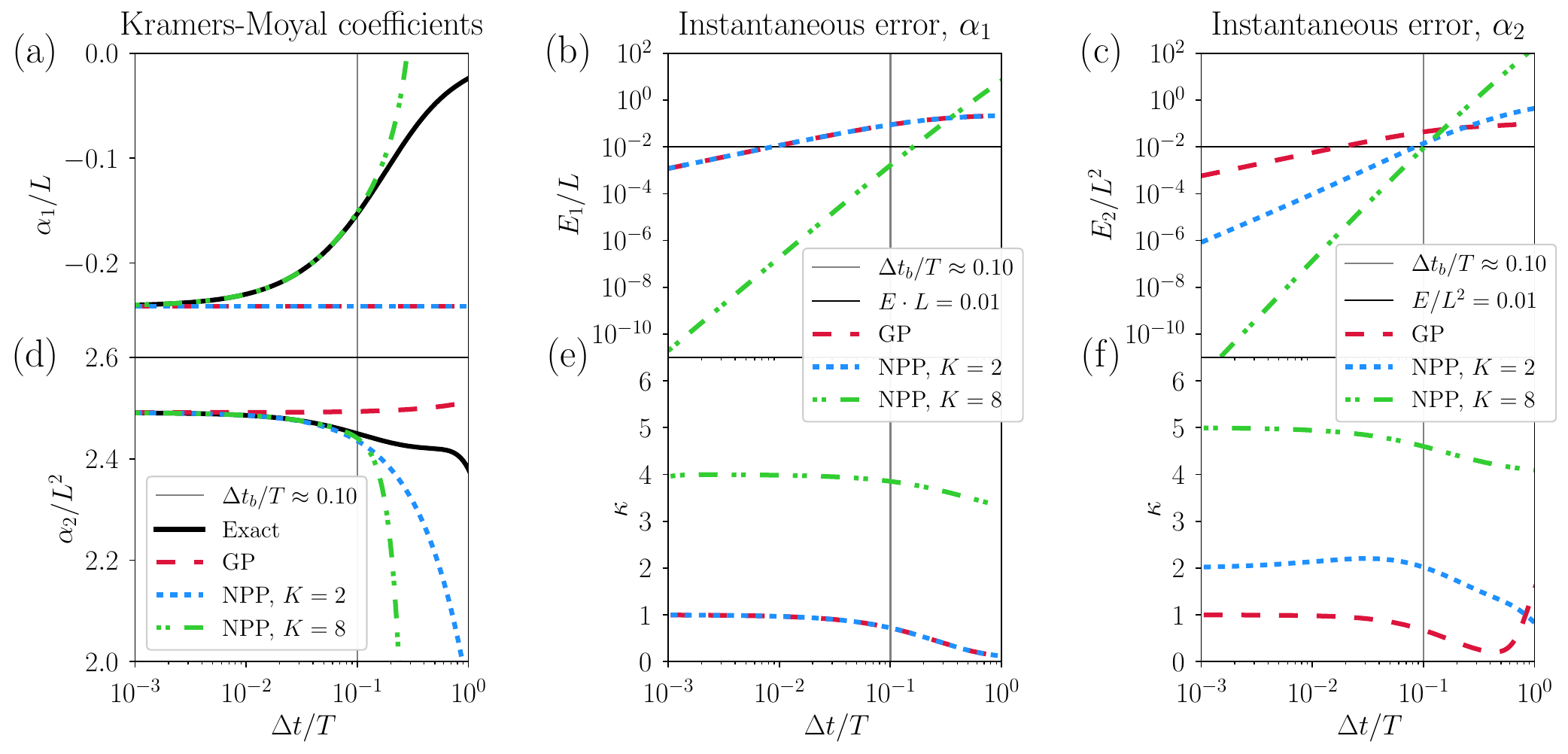} 
\caption{ \label{fig:moments} 
\textbf{Comparison of exact finite-time Kramers-Moyal coefficients with
 various approximations.}
Throughout this figure, data pertaining to the Gaussian 
propagator (GP) Eq.~\eqref{eq:langevin_propagator_final} is plotted as 
red dashed line;
data pertaining to the normalization-preserving propagator (NPP) 
Eq.~\eqref{eq:perturbative_solution_K}
with  $K=2$ is shown as 
blue dotted line,
and with $K=8$ as 
green dash-dotted line.
We show the (a) first and (d) second exact finite-time Kramers-Moyal coefficients as black solid lines,
together with various approximations.
The legend in (d) is also valid for (a).
While in (b), (c) we show the
instantaneous error Eq.~\eqref{eq:Error_KM}
 for the approximate finite-time Kramers-Moyal coefficients as a function of 
 the lagtime $\Delta t/T$,
 in (e), (f) we plot the corresponding running exponents Eq.~\eqref{eq:running_exponent}.
In all subplots, the vertical solid line indicates the breakdown
time $\Delta t_b/T \approx 0.10$ defined in Eq.~\eqref{eq:temporal_breakdown_condition}.
}
\end{figure*}

In Fig.~\ref{fig:comparison_various_orders}, we compare the exact propagator (black solid line)
with both the GP Eq.~\eqref{eq:langevin_propagator_final} (red dashed line),
and our NPP Eq.~\eqref{eq:perturbative_solution_K}
to order $\epsilon^2 \sim \Delta t$ ($K=2$, 
blue dotted line) and $\epsilon^8 \sim \Delta t^4$ ($K=8$, green dash-dotted line).

In Fig.~\ref{fig:comparison_various_orders} (a) we show the three
 perturbative solutions together 
with the exact analytical solution for $\Delta t/T = 0.05$, %(vertical gray dashed
i.e.~for a time shorter
than the breakdown time $\Delta t_b$.
We observe %in Fig.~\ref{fig:comparison_various_orders} (a)
 that for $\Delta t/T = 0.05$ all
perturbative solutions 
agree well with the analytical solution on the scales
used for the plot.
In Fig.~ \ref{fig:comparison_various_orders} (d) we show 
 the pointwise error Eq.~\eqref{eq:pointwise_error}
of the three perturbative solutions.
Consistent with subplot (a) we see that all pointwise errors are small compared 
to the typical values of the densities.
On the other hand, the pointwise error for the GP is almost one order of magnitude
 larger than the pointwise error of the
 NPP with $K=2$. %at order $\epsDL^2 \sim \Delta t$.
The pointwise errors of both the GP and the NPP with $K=2$ are
significantly larger than 
the pointwise error in the NPP with $K=8$,
which on the scales used for Fig.~\ref{fig:comparison_various_orders} (d) 
seems insignificant.
Consistent with these observations, the integrated pointwise errors 
Eq.~\eqref{eq:L1_error} are given by
$E_{\mathrm{GP}} \approx 0.0531$ (GP),
$E_{\mathrm{NPP}} \approx 0.0088$ (NPP, $K=2$),
and $E_{\mathrm{NPP}} \approx 0.0006$ (NPP, $K=8$).

In Fig.~\ref{fig:comparison_various_orders} (b) we plot the
perturbative and exact propagator for
$\Delta t= 0.2 T$, which is approximately twice the breakdown time $\Delta t_b$. 
We observe that now all approximate propagators deviate visibly from the exact
analytical solution.
While the GP deviates at the center of the distribution ($x/L \approx 0$), the 
deviations in the NPPs occur at the tails ($x/L \approx \pm 2$).
We confirm this in Fig.~\ref{fig:comparison_various_orders} (e),
where we show the pointwise error 
Eq.~\eqref{eq:pointwise_error}
for $\Delta t/T = 0.2$. 
We see that now the maximal error occurs in 
the NPP with $K=8$ %at order $\epsDL^8$ at
at
around $x/L \approx \pm 2$, 
where  we observe oscillations in the solution in Fig.~\ref{fig:comparison_various_orders} (b).
The corresponding integrated pointwise errors Eq.~\eqref{eq:L1_error},
evaluated at $\Delta t/T = 0.2$, $x_0/L = 0.5$, follow as
$E_{\mathrm{GP}} \approx 0.06$ (GP)
$E_{\mathrm{NPP}} \approx 0.05$ (NPP, $K=2$),
and $E_{\mathrm{NPP}} \approx 0.19$ (NPP, $K=8$).

In Fig.~\ref{fig:comparison_various_orders} (c) we show 
the instantaneous $L^1$-error Eq.~\eqref{eq:L1_error}
 of the perturbative propagators as a function
of the time increment $\Delta t$;
we indicate
the values $\Delta t/T = 0.05$, $0.2$ from the left and middle
 column of Fig.~\ref{fig:comparison_various_orders} 
as vertical broken lines, 
and the breakdown time $\Delta t_b/T \approx 0.10$ as a vertical solid line.
We observe
 that for the smallest lagtime considered, $\Delta t/T = 10^{-3}$,
the error in the GP Eq.~\eqref{eq:langevin_propagator_final}
is about five orders of magnitude larger than the error
of the NPP at order $\epsDL^2 \sim \Delta t$,
and about nine orders of magnitude larger than the error
of the NPP at order $\epsDL^8 \sim \Delta t^4$.
As the lagtime $\Delta t$ is increased,
 all errors display their respective expected power law scaling
\begin{equation}
E(\Delta t, x_0) \sim \Delta t^{\gamma+1/2},
\end{equation}
where $\gamma$ is the order of accuracy for the solution;
while for the GP Eq.~\eqref{eq:langevin_propagator_final} this is $\gamma = 0$, 
for the NPP
Eq.~\eqref{eq:perturbative_solution_rewritten_0} with 
order of the perturbation theory $K$ we have 
 $\gamma = K/2$.
To make the scalings more explicit we 
in Fig.~\ref{fig:comparison_various_orders} (f) plot the running exponent
\begin{align}
\label{eq:running_exponent}
\kappa(\Delta t) &\equiv \frac{ \partial \ln( E )}{\partial \ln(\Delta t) },
\end{align}
which is defined such that 
 if $E \sim \Delta t^{\alpha}$ for some real number $\alpha$, it holds that $\kappa =  \alpha$.
Figure ~\ref{fig:comparison_various_orders} (f) vividly demonstrates all anticipated power law scalings
of the error
 for small $\Delta t \ll \Delta t_b$.
As the time increment is increased to $\Delta t \approx \Delta t_b$, all errors 
start to deviate from their
respective power law scaling and, as Fig.~\ref{fig:comparison_various_orders} (c) shows,
 become of the order $E \cdot L \approx 0.01 = 1\%$.
For times larger than the breakdown time, the
NPP at order $\epsDL^8 \sim \Delta t^4$
starts to perform worse as compared to the lower-order propagators;
this is consistent with Fig.~\ref{fig:comparison_various_orders} (e).
We speculate that the higher-order NPP performs worse beyond 
the breakdown of perturbation theory because of 
 the higher polynomial order of the
 $\Qindex{k}$ (the polynomial order  increases as $3 k$);
 these high-order polynomials may result in very uncontrolled
 behavior. % beyond the breakdown of perturbation theory.

To summarize, Fig.~\ref{fig:comparison_various_orders} shows that
for small enough lagtime the 
NPP with $K=8$ 
significantly outperforms 
both the NPP with $K=2$,
as well as the GP Eq.~\eqref{eq:langevin_propagator_final}.
As the lagtime approaches the breakdown time $\Delta t_b$, 
all propagators approximate the exact analytical solution comparably well.
For lagtimes $\Delta t > \Delta t_b$,
all propagators yield an error $E \cdot L  > 0.01$, with the
NPP with $K=8$  producing the largest error out 
of the three propagators considered.

\subsection{Finite-time Kramers-Moyal coefficients}

We now consider the first two finite-time Kramers-Moyal
coefficients
Eqs.~\eqref{eq:alpha_one}, \eqref{eq:alpha_two} for our
  example system.

In Fig.~\ref{fig:moments} (a) we compare the exact first finite-time Kramers-Moyal coefficient
as a function of the lagtime (black solid line) 
to approximate evaluations, based on 
the GP Eq.~\eqref{eq:langevin_propagator_final} (red dashed line),
as well as the NPP Eq.~\eqref{eq:perturbative_solution_K} at order
 $\epsDL^2 \sim \Delta t$ ($K=2$, blue dotted line) and 
 $\epsDL^8\sim \Delta t^4$ ($K=8$, green dash-dotted line).
We observe that the GP and the NPP with $K=2$
lead to identical time-independent predictions, which start to deviate
from the exact result at the smallest lagtimes considered, $\Delta t/T \approx 10^{-3}$.
The NPP with $K=8$ on the other hand
 describes the exact result well for lagtimes up to $\Delta t \approx \Delta t_b \approx 0.10 T$.
 
 To quantify the deviations between exact and perturbative Kramers-Moyal coefficients, 
we consider the instantaneous error
 \begin{align}
 \label{eq:Error_KM}
 E_{i}(\Delta t,x_0) &= \left|
 \alpha_i(\Delta t,x_0) - \alpha_i^{\mathrm{e}}(\Delta t,x_0)
 %{\alpha_i^{\mathrm{e}}(\Delta t,x_0)}
  \right|,
 \end{align}
where $\alpha_i$ is the respective perturbative finite-time Kramers-Moyal
coefficient, and $\alpha_i^{\mathrm{e}}$ is the corresponding exact result.
In Fig.~\ref{fig:moments} (b) we plot the instantaneous error for the first
finite-time Kramers-Moyal
coefficient, for which $i=1$ in Eq.~\eqref{eq:Error_KM}.
In subplot (e) we show the corresponding
local exponent Eq.~\eqref{eq:running_exponent}.
Consistent with subplot (a), we in subplot (b) see that
 the instantaneous errors for GP and NPP with $K=2$
are identical.
According to Fig.~\ref{fig:moments} (e), 
 both errors scale as $E_{1} \sim \Delta t$,
 which confirms our error estimates Eqs.~\eqref{eq:EPP_one}, 
\eqref{eq:GP_KM}.
Figure \ref{fig:moments} (b) shows that 
for $\Delta t \lesssim \Delta t_b$, 
the error of
the NPP with $K = 8$ 
is typically orders of magnitude 
smaller than the errors of
 the GP and of the NPP with $K=2$.
In Fig.~\ref{fig:moments} (e) we observe a local exponent $\kappa = 4$ for the NPP 
with $K=8$, which is in agreement with Eq.~\eqref{eq:EPP_one}.

In Fig.~\ref{fig:moments} (d) we show plots of the second finite-time Kramers-Moyal
coefficient Eq.~\eqref{eq:alpha_two}.
Contrary to subplot (a), in subplot (d) the GP and NPP with $K=2$ are not identical.
While the GP starts to visibly deviate from the exact
 result at $\Delta t/T \approx 0.01$,
 the NPP with $K=2$ describes the exact result on the plotting scales up to 
 $\Delta t/T \approx 0.1$.
 At around the same lagtime, also 
the NPP with $K=8$ starts to visibly deviate from the exact result.
In  Fig.~\ref{fig:moments} (c) we show the 
instantaneous error Eq.~\eqref{eq:Error_KM} for $i = 2$.
We see that all approximate
 propagators lead to different power law scalings;
according to  Fig.~\ref{fig:moments} (f)
the respective power law exponents
of the errors are $\kappa= 1$ (GP), $\kappa = 2$ (NPP, $K=2$),
and  $\kappa= 5$ (NPP, $K=8$). 
These scalings are all consistent with our estimates 
Eqs.~\eqref{eq:EPP_two}, \eqref{eq:GP_KM}.

In summary, Fig.~\ref{fig:moments} shows
that both the GP and the NPP with $K=2$ predict the 
first two finite-time Kramers-Moyal coefficients only up to 
lagtimes that are significantly smaller than $\Delta t_b$. 
In comparison, Fig.~\ref{fig:moments} (b), (c) shows that
the pointwise error our NPP with $K=8$ 
is typically orders of magnitude smaller, and 
for $\Delta t < \Delta t_b$ is always less than $1\%$.
All power laws observed in Fig.~\ref{fig:moments}
are in agreement with our error
 estimates from Sect.~\ref{eq:error_estimates_observables}.

\subsection{Medium entropy production rate}

\begin{figure}[ht]
\centering \includegraphics[width=1\columnwidth]{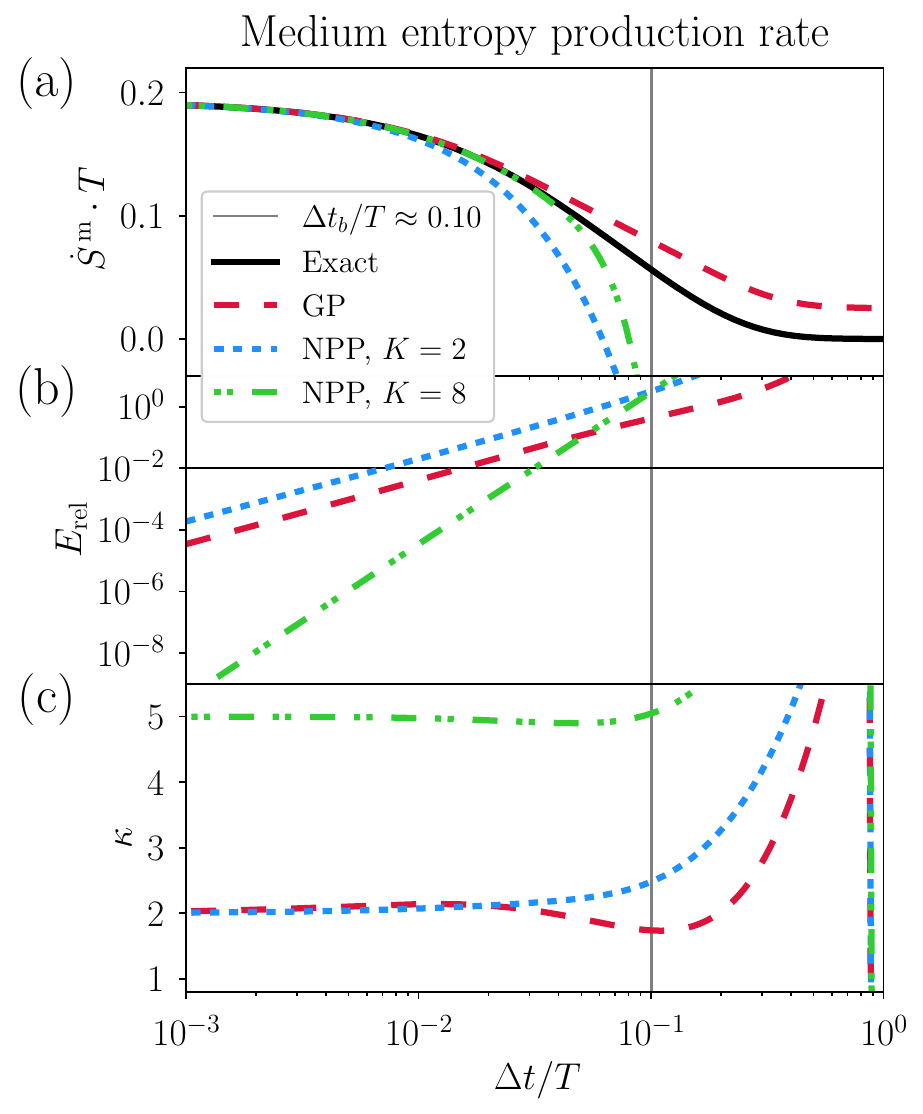} 
\caption{ \label{fig:medium_entropy_production} 
\textbf{Comparison of exact medium entropy production with approximations.}
Throughout this figure, we plot data pertaining to
 the GP Eq.~\eqref{eq:langevin_propagator_final} as 
red dashed line,
to the normalization-preserving propagator (NPP) 
Eq.~\eqref{eq:perturbative_solution_K}
with  $K=2$ as
blue dotted line,
and with $K=8$ as 
green  dash-dotted line.
The legend in (a) is also valid for (b), (c).
(a) We show the exact medium entropy production rate Eq.~\eqref{eq:def_Sm}
as a function of the lagtime $\Delta t$ as black solid line,
together with various approximation. 
While in (b) we show the instantaneous relative error Eq.~\eqref{eq:relative_error}
of the approximations from (a), we in (c)
show the running exponent Eq.~\eqref{eq:running_exponent}
of the relative error.
In all subplots, the vertical solid line indicates the breakdown
time $\Delta t_b/T \approx 0.10$ defined in Eq.~\eqref{eq:temporal_breakdown_condition}.
}
\end{figure}

We now turn to the medium entropy production rate Eq.~\eqref{eq:def_Sm}.
In Fig.~\ref{fig:medium_entropy_production} (a) we compare the exact medium entropy production
rate (black solid line) with approximations based on the
GP Eq.~\eqref{eq:langevin_propagator_final} (red dashed line), 
as well as the 
 NPP at order $\epsDL^2 \sim \Delta t$ ($K=2$, blue dotted line)
and at order $\epsDL^8 \sim \Delta t^4$ ($K = 8$, green dash-dotted line).
We see that for $\Delta t/T < 10^{-2}$, all approximations describe the exact
results very well on the plotting scales. 
We confirm this in Fig.~\ref{fig:medium_entropy_production} (b), where we plot the instantaneous relative error
\begin{align}
\label{eq:relative_error}
E_{\mathrm{rel}}(\Delta t)  &= \left| \frac{ \dot{S}^{\mathrm{m},\mathrm{e}}(\Delta t) -  \dot{S}^{\mathrm{m}}(\Delta t)}{\dot{S}^{\mathrm{m},\mathrm{e}}(\Delta t)} \right|,
\end{align}
where here $\dot{S}^{\mathrm{m},\mathrm{e}}$, $\dot{S}^{\mathrm{m}}$ stand 
for the exact and approximate medium entropy production, respectively.
From Fig.~\ref{fig:medium_entropy_production} (b) we observe that while 
the NPP approximation with $K=2$
surpasses 1\% relative error at $\Delta t/T \approx 7 \times 10^{-3}$,
the GP does so at $\Delta t/ T \approx 1.5\cdot 10^{-2}$, and the
the $K=8$ approximation at $\Delta t /T \approx 3.1 \cdot 10^{-2}$.
Finally, in Fig.~\ref{fig:medium_entropy_production} (c) we show the 
running exponent Eq.~\eqref{eq:running_exponent} pertaining to the 
relative error Eq.~\eqref{eq:relative_error}.
As expected, before the breakdown time $\Delta t_b$, 
the error for the GP and the NPP with $K=2$
scales as $\Delta t^2$, which for the GP is in agreement with Eqs.~\eqref{eq:error_GP_f}.
Similarly, the error for the NPP with $K=8$ scales as $\Delta t^5$.
Both error scalings for the NPP are anticipated, as we evaluate the perturbative medium 
entropy production to orders $\Delta t$ (corresponding to $K=2$) and $\Delta t^4$ 
(corresponding to $K=8$).

\section{Summary and conclusions}
\label{sec:conclusions}

In this work we present a perturbation approach for evaluating the short-time 
Fokker-Planck propagator to in principle arbitrary precision.
We provide two representations of the resulting perturbative propagator,
namely the normalization-preserving propagator (NPP)
and the positivity-preserving propagator (PPP) \cite{ait-sahalia_maximum_2002,
ait-sahalia_closed-form_2008}.
The NPP preserves the proper normalization of
the propagator exactly, but can take on negative values outside of its 
regime of validity, i.e.~for 
very improbable large increments $\Delta x = x - x_0$,
or for
time increments $\Delta t =t  -t_0$ so large that our perturbation ansatz breaks down.
The PPP on the other hand is manifestly positive, but in general will not preserve the proper
normalization of a probability density; more so, depending on the diffusivity and drift profiles,
 the PPP might not even fulfill the proper boundary conditions of the 
 underlying Fokker-Planck equation (FPE).
Since it allows for straightforward calculation of expectation values, we use the
NPP to evaluate perturbative expressions for the moments $\langle \Delta x^n \rangle$,
the first two finite-time Kramers-Moyal coefficients,
as well as the medium entropy production.
We derive error estimates for both our perturbative propagators and  expectation values,
and compare our results to the corresponding errors of the standard Gaussian short-time propagator (GP).
Remarkably, we find that the
GP
in general has an integrated pointwise absolute error of the order of $\sqrt{\Delta t}$, i.e.~is a sublinear
approximation to the true short-time propagator;
still, the GP allows to evaluate expectation values with oder-$\Delta t$ accuracy.
We illustrate all our results via an explicit numerical example,
where we find that our approximation propagators can outperform the GP in terms of accuracy
by orders of magnitude. 
However, in regimes where the perturbation theory breaks down (e.g.~too large lagtimes),
the error of our approximations can become worse than that of the GP;
we speculate that this is due to the high-order polynomials that our perturbation theory contains.

Accurate approximations of the short-time propagator
and of short-time expectation values have several  applications
that are fundamental to the practical use of diffusive stochastic dynamics models.
One such application is the parametrization of the FPE Eq.~\eqref{eq:fokker_planck}
from observed time series.
For this, one measures expectation values of short-time observables,
such as the first two finite-time Kramers-Moyal coefficients 
$\langle \Delta x \rangle /\Delta t$,
$\langle \Delta x^2 \rangle /\Delta t$.
These coefficients are expressed in terms of drift and diffusivity via
theoretical estimates such as Eqs.~\eqref{eq:alpha_one}, \eqref{eq:alpha_two},
or higher-order versions thereof (which we provide in our python module \cite{PySTFP}).
The FPE
is then parametrized by choosing 
 drift and diffusivity such that the theoretical estimates reproduce the
 measured expectation values  \cite{gladrow_experimental_2019,
gladrow_experimental_2021,
thorneywork_resolution_2024}.
Here, more accurate theoretical estimates of the finite-time Kramers-Moyal coefficients
 reduce one source of error, and will be in particular relevant for time series that can only
 be observed with relatively low temporal frequency.
 The same holds true for other methods of parameter fitting, such 
 as the maximum likelihood approach for observed transitions \cite{dacunha-castelle_estimation_1986}.

A related application of our short-time propagator is the numerical sampling of time-discretized 
path-spaces via Markov Chain Monte Carlo (MCMC), \cite{gilks_markov_1995,
tse_estimation_2004,
beskos_mcmc_2008,
lin_generating_2010,
durmus_fast_2016,
pieschner_bayesian_2020}.
This can be used for calculating expectation values over path ensembles,
or for sampling the posterior distribution
 in Bayesian parameter estimation or augmentation \cite{dacunha-castelle_estimation_1986,
elerian_note_1998,
tse_estimation_2004,
pieschner_bayesian_2020}.
Here, a more accurate short-time propagator allows to use
 larger time increments $\Delta t$ for the time-slicing discretization of the probability density on 
 path space, 
 which results in a lower-dimensional sampling problem.

 Another possible avenue for further research is the relation between 
our PPP Eq.~\eqref{eq:short_time_propagator_exp_dt}
and the multiplicative-noise stochastic actions that have been derived in 
the literature \cite{arnold_langevin_2000,
tang_summing_2014,
chow_path_2015,
cugliandolo_rules_2017,
cugliandolo_building_2019,
moreno_conditional_2019,
de_pirey_path_2023}.
For the special case of additive noise, where the diffusivity is independent 
of $x$, the PPP 
was recently used to 
discuss that the form of the path-integral action
depends on whether one evaluates the action on a differentiable path
 (where increments scale as $\Delta x \sim \Delta t$), or a
typical realization of the It\^{o}-Langevin Eq.~\eqref{eq:LangevinEq} 
(where increments scale as $\Delta x \sim \sqrt{ \Delta t}$) 
\cite{gladrow_experimental_2021}.
For this analysis, it was critical to consider the short-time propagator to order $\Delta t$ 
in the exponent, and so in Ref.~\cite{gladrow_experimental_2021} an approximation
 that goes beyond the usual Gaussian propagator Eq.~\eqref{eq:langevin_propagator_final} 
was employed.
For multiplicative noise, defining and working with a path-integral action is in general
more intricate \cite{durr_onsager-machlup_1978,
cugliandolo_rules_2017,
cugliandolo_building_2019,
moreno_conditional_2019,
de_pirey_path_2023,
kappler_measurement_2022,
thorneywork_resolution_2024}.
Similarly to the result of Ref.~\cite{gladrow_experimental_2021} discussed above,
it will be interesting to investigate whether also
for the multiplicative-noise stochastic action,
our %multiplicative-noise 
PPP to order $\Delta t$
(which we give explicitly to second order in perturbation theory in 
App.~\ref{sec:physical_units}),
will lead to insights that cannot be obtained from the Gaussian propagator
Eq.~\eqref{eq:langevin_propagator_final}.

We here derive the short-time propagator only for a one-dimensional 
reaction coordinate.
It is an open question whether derivation can be generalized also to multidimensional systems,
for which one would perturb around a multivariate Gaussian distribution,
as explored in Refs.~\cite{ait-sahalia_closed-form_2008,sorkin_consistent_2024} by different means.
The key question here will be whether the multidimensional equivalent of
Eq.~\eqref{eq:dimensionless_hierarchy_Q}
allows for systematic solution.

\acknowledgements{We thank Ronojoy Adhikari, 
Michael E.~Cates, and Erwin Frey for many helpful discussions.
Work funded in part by the European Research Council under the Horizon 2020 Programme, 
ERC grant agreement number 740269.
We acknowledge funding from the
 European Union's Horizon 2020 research and innovation programme 
 under the Marie Sk\l{}odowska-Curie grant agreement No 101068745.}

\appendix 

\section{Derivation and properties of the perturbative solution}
\label{app:derivation}

\subsection{Solution of the recursive equation}
\label{app:solution_recursive}

We now
derive a recursive scheme to solve
 Eq.~\eqref{eq:dimensionless_hierarchy_Q}, 
 and show 
 that the resulting solution $\Qindex{k}(\xDL)$ at order $k$ is a polynomial in $\xDL$
of order at most $3k$.
We proceed by induction. 

For the base case $k =0$ we note that $\Qindex{0} = 1$ 
solves Eq.~\eqref{eq:dimensionless_hierarchy_Q},
and is a 
polynomial in $\xDL$ of order $3k = 0$

For the induction step we 
assume that for a given $k \geq 1$, 
each $\Qindex{l}$ with $l <k$ is a
known polynomial in $\xDL$
of order at most $3l$.
We can then evaluate 
the right-hand side of Eq.~\eqref{eq:dimensionless_hierarchy_Q}, 
which results in a polynomial in $\xDL$.
By counting the highest powers of each term on the right-hand side 
of Eq.~\eqref{eq:dimensionless_hierarchy_Q},
we see that this polynomial is of order at most $3k$.
More explicitly, 
the
 highest possible power in $\xDL$ is obtained from 
the $l=1$ term of the second sum on the right-hand side of
Eq.~\eqref{eq:dimensionless_hierarchy_Q},
and more specifically by the term
$\xDL^{l+2} \Qindex{k-l}$
in Eq.~\eqref{eq:operator_L_D}:
For $l=1$, and using that $\Qindex{k-l}$ is at
most of order $\xDL^{3(k-l)}$, we see that 
the highest possible power in $\xDL^{l+2} \Qindex{k-l}$
is $\xDL^{l+2} \xDL^{3(k-l)} = \xDL^{3k}$.

We now have established that in the induction step, the right-hand side of 
the inhomogeneous differential Eq.~\eqref{eq:dimensionless_hierarchy_Q} 
is a polynomial
of order at most $\xDL^{3k}$.
Since the differential operator on the left-hand side of 
Eq.~\eqref{eq:dimensionless_hierarchy_Q} is linear, we can obtain a
solution for the polynomial inhomogeneity 
by summing over 
for all the possible monomials 
$\xDL^0$, $\xDL^1$, ..., $\xDL^{3k}$ that form the polynomial.

For a monomial inhomogeneity $\xDL^q$ with $q \in \mathbb{N}_0$, Eq.~\eqref{eq:dimensionless_hierarchy_Q} reads
 \begin{align}
\label{eq:dimensionless_hierarchy_inhom_monomial}
\partial_{\xDL}^2 &f
- \xDL \partial_{\xDL} f
- k f
= \xDL^q,
\end{align}
and by direct substitution it follows that a solution to this equation is
\begin{equation}
\label{eq:dimensionless_hierarchy_inhom_monomial_solution}
f(\xDL) = -\frac{1}{k+q}\sum_{n = 0}^{\left \lfloor{q/2}\right \rfloor }
 \xDL^{q-2n}
\prod_{i=0}^{n-1}\frac{(q-2i)(q-1-2i)}{k+q-2(i+1)},
\end{equation}
with the floor function
\begin{equation}
\left \lfloor{q/2}\right \rfloor = \begin{cases}
q/2 & \mathrm{if}~q~\mathrm{even},\\
(q-1)/2 & \mathrm{if}~q~\mathrm{odd}.
\end{cases}
\end{equation}
We see that the solution Eq.~\eqref{eq:dimensionless_hierarchy_inhom_monomial_solution} 
is a polynomial in $\xDL$, and is of the same order as
the monomial inhomogeneity on the right-hand side of 
Eq.~\eqref{eq:dimensionless_hierarchy_inhom_monomial}. 

This finally implies that for the polynomial of order at most $3k$ on the right-hand side of
Eq.~\eqref{eq:dimensionless_hierarchy_Q}, we have a solution $\Qindex{k}$
that is a linear combination
 of terms of the form Eq.~\eqref{eq:dimensionless_hierarchy_inhom_monomial_solution},
and as such is a polynomial in $\xDL$ of order at most $3k$, as claimed.
We now show that this solution fulfills both the boundary and normalization condition.

First, 
for our polynomial solution $\Qindex{k}(\xDL)$ 
 the product
\begin{equation}
\PDL^{(k)}(\xDL) \equiv \Qindex{k}(\xDL) \PDL^{(0)}(\xDL)
\label{eq:PDL_ansatz_Q}
\end{equation}
fulfills the boundary conditions
 Eq.~\eqref{eq:PDL_power_series_boundary_condition},
 because a product of an exponential decay of $P^{(0)}$ with a finite polynomial 
 eventually always
 decays.
Our algorithm
 therefore leads to a term $\PDL^{(k)}(\xDL)$ which solve the second-order ODE
 Eq.~\eqref{eq:dimensionless_hierarchy}
 with the boundary conditions Eq.~\eqref{eq:PDL_power_series_boundary_condition} at $\xDL = \pm \infty$.

To show that the perturbative solution Eq.~\eqref{eq:perturbative_solution_rewritten_0} 
conserves probability exactly, we rewrite
the power series as
\begin{align}
\label{eq:PDL_power_series}
\PDL(\xDL,\tDL) &= \sum_{k=0}^{\infty} \epsDL^k(\tDL) \PDL^{(k)}(\xDL),
\end{align}
where $P^{(k)}$ is defined in Eq.~\eqref{eq:PDL_ansatz_Q}.
Substituting Eqs.~\eqref{eq:PDL_power_series}, 
\eqref{eq:aDL_power_series}, \eqref{eq:DDL_power_series} into 
Eq.~\eqref{eq:fokker_planck_dimensionless}, 
and demanding that the resulting equation hold at each order
in $\epsDL$ separately, we obtain 
\begin{align}
\label{eq:dimensionless_hierarchy}
\partial_{\xDL}^2 \PDL^{(k)} &
+ \partial_{\xDL} \left( \xDL \PDL^{(k)} \right)
- k \PDL^{(k)}
\\ 
&= 
\sum_{l=0}^{k-1}
\Aindex{l}
\partial_{\xDL} 
\left[ 
	\xDL^{l} \PDL^{(k-1-l)} 
\right]
- 
\sum_{l=1}^{k}
\Dindex{l}
\partial_{\xDL}^2
\left[ 
\xDL^l
\PDL^{k-l}
\right],
\nonumber
\end{align}
where the sums on the right-hand side are defined as zero if the upper summation 
bound is smaller than the lower summation bound.
Note that upon substituting Eq.~\eqref{eq:PDL_ansatz_Q}
into Eq.~\eqref{eq:dimensionless_hierarchy}
we recover Eq.~\eqref{eq:dimensionless_hierarchy_Q}.

By integrating Eq.~\eqref{eq:dimensionless_hierarchy} over $\xDL$ from $-\infty$ to
$\infty$, and
noting that from the form of Eq.~\eqref{eq:PDL_ansatz_Q} with $\PDL^{(0)}$ Gaussian and $\Qindex{k}$
 polynomial it follows that
 all terms involving spatial derivatives of $\PDL^{(k)}$
vanish for $|\xDL| \rightarrow \infty$,
we see that for $k \geq 1$ it holds that
\begin{align}
\int_{-\infty}^{\infty}\mathrm{d}\xDL\,\PDL^{(k)}(\xDL) &= 0,
\end{align}
which implies 
\begin{align}
\int_{-\infty}^{\infty}\mathrm{d}\xDL\,\PDL(\xDL,\tDL) 
&= 
\sum_{k=0}^{\infty} \epsDL^k(\tDL) \int_{-\infty}^{\infty}\mathrm{d}\xDL\,\PDL^{(k)}(\xDL)
\\
&= 
 \int_{-\infty}^{\infty}\mathrm{d}\xDL\,\PDL^{(0)}(\xDL) = 1.
\end{align}
This shows that the perturbation series Eq.~\eqref{eq:perturbative_solution_rewritten_0}
 conserves probability exactly at all orders.

\subsection{Parity of the polynomials $\Qindex{k}$}
\label{app:parity}

We now show that the polynomial $\Qindex{k}(\xDL)$
is comprised of only even powers of $\xDL$ for $k$ even (and $k=0$) and
only of odd powers of $\xDL$ for $k$ odd.
Phrased differently, 
we show 
that for all $k \in \mathbb{N}_0$ we have
\begin{align}
\label{eq:Q_parity}
\Qindex{k}(\xDL) &= (-1)^{k} \Qindex{k}(-\xDL).
\end{align}

We proceed by induction in the order $k$. 
For $k=0$ it follows from Eq.~\eqref{eq:Qindex0} that 
$\Qindex{0}$ fulfills Eq.~\eqref{eq:Q_parity}.
For the induction step, we assume $k \geq 1$ and that
 all $\Qindex{l}$ fulfill Eq.~\eqref{eq:Q_parity} for $l < k$.
Then according to Eqs.~\eqref{eq:operator_L_A}, \eqref{eq:operator_L_D},
the right-hand side of Eq.~\eqref{eq:dimensionless_hierarchy_Q} is an
 even polynomial 
in $\xDL$ for $k$ even, and an odd polynomial in $\xDL$ for $k$ odd.
From Eq.~\eqref{eq:dimensionless_hierarchy_inhom_monomial_solution} it 
then follows that also $\Qindex{k}$ is an even (odd) polynomial for $k$ even (odd).

We note that this in particular implies that for $n+k$ odd it holds that
\begin{align}
\langle \xDL^n \rangle^{(k)} &\equiv \int_{-\infty}^{\infty}d\xDL\,\xDL^n \Qindex{k}(\xDL) \PDL^{(0)}(\xDL) = 0,
\end{align}
since $\PDL^{(0)}$
is an even function in $\xDL$ 
and the integral boundaries are symmetric around $\xDL = 0$.

\section{Second-order perturbative propagator in physical units}
\label{sec:physical_units}

In Sect.~\ref{eq:perturbative_solution_of_FP} of the main text
we discuss both the normalization-preserving propagator (NPP)
and the positivity-preserving propagator (PPP).
In the present appendix we first give explicit expressions for both to second
order in perturbation theory, i.e.~to order $\epsDL^2 \sim \Delta t$.
We then rewrite the PPP in terms of an arbitrary discretization scheme,
from which we
 recover the usual pathwise medium entropy production using a midpoint discretization scheme.
Finally, we validate the perturbative propagators from this appendix
by comparison to an exact solution, similar to Sect.~\ref{sec:numerical_example}.

\subsection{Normalization-preserving propagator}
\label{sec:NP_rep_dt}

Using the definitions of the dimensionless units Eqs.~(\ref{eq:tDL_def}-\ref{eq:def_DDL}),
we can cast the NPP Eq.~\eqref{eq:perturbative_solution_K} back into physical units. 
To order $\epsDL^2 \sim \Delta t$ in perturbation theory, this yields
\begin{align}
\label{eq:NPP_quadratic}
P(x,t \mid x_0, t_0) =& \frac{1}{\sqrt{4 \pi D(x_0) \Delta t}}\exp\left[ - \frac{ \Delta x^2}{4 D(x_0) \Delta t}\right]
\\
&  \times
\nonumber
\left[ 1+ \sqrt{ \Delta t}Q^{(1/2)} +\Delta t Q^{(1)} + \mathcal{O}(\Delta t^{3/2})\right]
\end{align}
with $Q^{(k)} \equiv Q^{(k)}(\Delta x, \Delta t) \equiv \epsDL^k(\tDL) \Qindex{k}(\xDL)/\sqrt{ \Delta t}^k$, so that
\begin{align}
Q^{(1/2)} &= 
\frac{\xDL}{2 \sqrt{D}}
 \left[ 2 a - 3 \partial_x D
+ \partial_xD
\xDL^2
\right]
\\
 Q^{(1)}&= 
2\left( \frac{ a^2}{9 D} + \frac{ \partial_x a}{4} \right)
\left(  \xDL^2 - 1\right)
\nonumber
+\frac{ a \partial_x D}{4 D} 
\left( \xDL^4 - 5 \xDL^2 + 2\right)
\\ & \qquad +
\frac{ (\partial_x D)^2}{ 16 D} 
\left( \xDL^6 - 11 \xDL^4 + 21 \xDL^2 - 3
\right)
\\ & \qquad +
\frac{ \partial_x^2 D}{12} \left( 2 \xDL^4 - 9 \xDL^2 + 3 \right),
\nonumber
\end{align}
where $D$, $a$ as well as their spatial derivatives are evaluated at $x_0$,
and where as before $\xDL \equiv \Delta x/R \equiv \Delta x/\sqrt{ 2 D(x_0)\Delta t}$.

\subsection{Positivity-preserving propagator}
\label{sec:PP_rep_dt}
To obtain the PPP Eq.~\eqref{eq:perturbative_solution_rewritten_4} to order $\Delta t$
in physical units,
we substitute Eqs.~\eqref{eq:Qindex0}, 
\eqref{eq:Qindex1}, \eqref{eq:Qindex2} 
into Eq.~\eqref{eq:perturbative_solution_rewritten_4},
and substitute the definitions of our dimensionless units.
This yields the short-time propagator 
\begin{align}
\label{eq:short_time_propagator_exp_dt}
P(x,t\mid x_0,t_0) &= \frac{1}{\sqrt{ 4 \pi D(x_0) \Delta t}} \exp\left( - \Delta S \right)
\end{align}
with
\begin{align}
\label{eq:S_series_exp}
\Delta S &=  
\Delta S^{(0)}
+ 
\sqrt{ \Delta t} \Delta S^{(1/2)}
+ 
\Delta t \Delta S^{(1)}
+ \mathcal{O}(\Delta t^{3/2}),
\end{align}
where
\begin{align}
\Delta S^{(0)} &= \frac{\Delta t}{4D}\left( \frac{\Delta x}{\Delta t} - a\right)^2, 
\\
\Delta S^{(1/2)} &=
\partial_x D \sqrt{\frac{2}{D}} \frac{ \xDL}{4} \left( 3 - \xDL^2\right)
\\
\label{eq:PP_S1}
\Delta S^{(1)} &=
\frac{\partial_x a}{2}\left( 1 - \xDL^2\right)
+ \frac{\partial_x^2 D}{12} \left( -2\xDL^4 + 9\xDL^2 - 3\right)
\\ & \qquad \nonumber
+ \frac{ (\partial_x D)^2}{16 D}\left( 5 \xDL^4 - 12 \xDL^2 + 3\right)
+ \frac{ a \partial_xD }{2D} \left( \xDL^2 - 1 \right)
\end{align}
 where  $D$, $a$ and their spatial derivatives are all evaluated at $x_0$,
 and where as before $\xDL \equiv \Delta x/R \equiv \Delta x/\sqrt{ 2 D(x_0) \Delta t}$.
Recalling that the typical leading-order scaling of the
 increment is $\Delta x = \mathcal{O}(\Delta t^{1/2})$
in the range where Eq.~\eqref{eq:short_time_propagator_exp_dt} is 
non-negligible,
we note that  $\Delta S^{(0)}$ contains terms of order $\Delta t^{0}$, $\Delta t^{1/2}$, $\Delta t^{1}$.
On the other hand, all terms of $\Delta S^{(1/2)}$ scale to leading order as $\Delta t^{1/2}$
and all terms of $\Delta S^{(1)}$ scale to leading order as $\Delta t$.

For additive noise all terms that include $\partial_x D$, $\partial_x^2 D$ vanish, and
Eq.~\eqref{eq:S_series_exp} reduces to the short-time propagator
 from Ref.~\cite{gladrow_experimental_2021}, 
which is given by
\begin{align}
\label{eq:DS_additive_noise}
\Delta S &= \frac{\Delta t}{4D}\left( \frac{\Delta x}{\Delta t} - a\right)^2
- 
\frac{\partial_x a}{2D} \left( \Delta x^2 - 2 D\Delta t \right)
+ \mathcal{O}(\Delta t^{3/2}).
\end{align}
The highest power in $\Delta x$ which appears in this expression is $\Delta x^2$, 
with a prefactor $1/(4D\Delta t) - \partial_x a /(2D)$.
For small enough  timestep $\Delta t$ this prefactor is positive,
so that in that case 
the resulting propagator Eq.~\eqref{eq:short_time_propagator_exp_dt} 
fulfills the boundary conditions Eq.~\eqref{eq:FP_BC_DL}.

To discuss the boundary conditions for the general case of multiplicative noise, 
we note that 
the highest power in $\xDL$ that appears in Eq.~\eqref{eq:S_series_exp}
is $\xDL^4$. According to Eq.~\eqref{eq:PP_S1}, the term with the highest power is given by
$\Delta t \xDL^4/2 [ -  \partial_x^2 D/3 + 5\partial_x D /(8D)]$.
If this factor has a negative sign, then the exponential in Eq.~\eqref{eq:short_time_propagator_exp_dt}
diverges as $|\xDL | \rightarrow \infty$.
Thus, the representation Eq.~\eqref{eq:short_time_propagator_exp_dt}
of the short-time propagator only fulfills the boundary conditions
Eq.~\eqref{eq:FP_BC_DL} 
if 
\begin{align}
\label{eq:condition_BC_exp}
\frac{15}{8}\frac{(\partial_x D)^2(x_0)}{D(x_0)} > (\partial_x^2 D)(x_0) .
\end{align}
This condition is independent of the temporal breakdown condition 
Eq.~\eqref{eq:temporal_breakdown_condition}.
We mathematically associate
Eq.~\eqref{eq:condition_BC_exp} with the finite radius of convergence $|z| = 1$
of the Taylor series
Eq.~\eqref{eq:ln_taylor} used in deriving the PPP;
as $|\xDL| \rightarrow \infty$, the corresponding values
 for $|z|$ are typically larger than $1$.
At this point we emphasize again that the short-time propagator should only 
be non-negligible for $|\xDL| \lesssim 1$ due to the finite speed of diffusion;
 the breakdown of the boundary conditions in the PPP hence highlights the
technical challenges for deriving 
 approximate short-time solutions
of the FP Eq.~\eqref{eq:fokker_planck}.

\subsection{Midpoint discretization and path-wise entropy production}
\label{sec:midpoint_discretization}

In all our results
so far,
$a$, $D$ and all their derivatives are evaluated at the initial point $x_0$,
which corresponds to an initial point discretization scheme.
To show how the propagator with respect to other discretization schemes
is obtained from our results, we now rewrite the PPP Eq.~\eqref{eq:short_time_propagator_exp_dt}
with respect to an arbitrary discretization scheme, defined by a parameter $\alpha \in [0,1]$.
For this, we note that 
\begin{equation}
x_0 = x_{\alpha} -  \alpha \Delta x,
\end{equation}
where $x_{\alpha} = x_0 +\alpha \Delta x$.
While for $\alpha = 0$ we have $x_{\alpha} = x_0$ (initial point discretization scheme),
for $\alpha = 1/2$ we have $x_{\alpha} = (x_0 + x)/2$ (midpoint discretization scheme).
The value of any analytical function, $f(x_0)$, can be expressed via Taylor expansion 
around $x_{\alpha}$ as
\begin{equation}
f(x_0) = f({x}_\alpha) 
- \alpha  \Delta x (\partial_x f)({x}_\alpha)
+\frac{\alpha^2}{2} \Delta x^2(\partial_x^2 f)({x}_\alpha) + \mathcal{O}(\Delta x^3)
\end{equation}
Substituting this Taylor expansion for $f \equiv a$, $D$, and their derivatives,
into Eq.~\eqref{eq:short_time_propagator_exp_dt}, we obtain
\begin{align}
\label{eq:short_time_propagator_exp_alpha_dt}
P(x,t\mid x_0,t_0) &= \frac{1}{\sqrt{ 4 \pi {D}_\alpha \Delta t}} \exp\left( - {\Delta {S}_\alpha} \right)
\end{align}
where
\begin{align}
\label{eq:delta_alpha_S}
\Delta {S}_\alpha &=  
\Delta {S}_\alpha^{(0)}
+ 
\sqrt{\Delta t}
\Delta {S}_\alpha^{(1/2)}
+ 
\Delta t
\Delta {S}_\alpha^{(1)}
+ \mathcal{O}(\Delta t^{3/2}),
\end{align}
with the functions $\Delta \bar{S}^{(k)} \equiv \Delta \bar{S}^{(k)}(\Delta x, \Delta t, \bar{X})$ 
are given by
\begin{align}
\label{eq:delta_alpha_S0}
\Delta {S}_\alpha^{(0)}
 &= \frac{\Delta t}{4{D}_\alpha}\left( \frac{\Delta x}{\Delta t} - {a}_\alpha + 2\alpha \partial_x {D}_\alpha\right)^2, 
\\
\label{eq:delta_alpha_Shalf}
\Delta {S}_\alpha^{(1/2)}
 &= 
 \partial_x D_{\alpha} \sqrt{ \frac{ 2}{D_\alpha}} \frac{ \xDLalpha}{4}
 \left( 2 \alpha - 1 \right) (\xDLalpha^2 - 3)
\\
\Delta {S}^{(1)}_\alpha &=
\frac{ \partial_x a_{\alpha}}{2} \left( 2 \alpha \xDLalpha^2 - \xDLalpha^2 + 1\right)
\nonumber
\\ & \quad + \nonumber
\frac{\partial_x^2 D_\alpha}{12}\left( 
-6 \alpha^2 \xDLalpha^2 (\xDLalpha^2 - 1)
\right. \\ & \qquad \qquad \left. \nonumber
+ 6 \alpha \xDLalpha^2(\xDLalpha^2 - 3)
- 2 \xDLalpha^4 + 9\xDLalpha^2 - 3
\right)
\\ & \quad +
\frac{ a_\alpha \partial_x D_\alpha}{2 D_\alpha}
\left( \xDLalpha^2 - 1\right)(1 -2 \alpha)
 \label{eq:delta_alpha_S1}
\\ & \quad +
\nonumber
\frac{ \left(\partial_x D_\alpha\right)^2}{16 D_\alpha}
\left( 
16 \alpha^2 \xDLalpha^4 
- 8 \alpha^2 \xDLalpha^2
- 16 \alpha^2
- 16 \alpha \xDLalpha^4
\right. \\ & \qquad \qquad \left.
+ 24 \alpha \xDLalpha^2
+ 5 \xDLalpha^4
- 12 \xDLalpha^2
+ 3
\right).
\nonumber
\end{align}
Here, $\xDLalpha \equiv \Delta x/\sqrt{2 D_\alpha \Delta t}$.

For the midpoint convention, $\alpha = 1/2$, we write
\begin{align}
\xDL_{\alpha=1/2} \equiv \bar{x} \equiv \frac{x_0 + x}{2},
\end{align}
and it holds that $\Delta {S}_\alpha^{(1/2)} = 0$.
We therefore 
 write Eq.~\eqref{eq:delta_alpha_S} as
\begin{align}
\label{eq:delta_bar_S}
\Delta \bar{S} &=  
\Delta \bar{S}^{(0)}
+ 
\Delta t \Delta \bar{S}^{(1)}
+ \mathcal{O}(\Delta t^{3/2}),
\end{align}
with the functions $\Delta \bar{S}^{(k)} \equiv \Delta \bar{S}^{(k)}(\Delta x, \Delta t, \bar{X})$ 
from Eqs.~\eqref{eq:delta_alpha_S0}, \eqref{eq:delta_alpha_S1} reducing to 
\begin{align}
\label{eq:delta_bar_S0} 
\Delta \bar{S}^{(0)}
 &= \frac{\Delta t}{4\bar{D}}\left( \frac{\Delta x}{\Delta t} - \bar{a} + \partial_x \bar{D}\right)^2, 
\\
\nonumber
\Delta \bar{S}^{(1)} &=
\frac{ 1}{2} \partial_x \bar{a}
+ \frac{ \partial_x^2 \bar{D}}{24} \left( - \bar{\xDL}^4 + 3 \bar{\xDL}^2 - 6 \right)
\\
& \quad + 
\frac{ (\partial_x \bar{D})^2}{16 \bar{D}} 
\left( 
\bar{\xDL}^4 - 2 \bar{\xDL}^2 - 1
\right).
 \label{eq:delta_bar_S1}
\end{align}
We here use a bar to denote that the expression is evaluated at $\bar{x}$, 
e.g.~$\bar{a} \equiv a(\bar{x})$, $\partial_x \bar{D} \equiv (\partial_x D)(\bar{x})$,
and where $\bar{\xDL} \equiv \Delta x/\sqrt{ 2 \bar{D}\Delta t}$.
We remark that despite the lack of an explicit term $\Delta \bar{S}^{(1/2)}$ in Eq.~\eqref{eq:delta_bar_S}, 
the contribution Eq.~\eqref{eq:delta_bar_S0} contains terms of order $\Delta x \sim \sqrt{ \Delta t}$.

As for the PPP with initial-point evaluation, Eq.~\eqref{eq:short_time_propagator_exp_dt}, 
there is no a-priori guarantee that the midpoint-evaluated propagator with 
Eq.~\eqref{eq:delta_bar_S} decays to
 zero as $|\Delta x | \rightarrow \infty$.
The leading order term in $\Delta x$ is given by the quartic powers in Eq.~\eqref{eq:delta_bar_S1},
so that for large $|\Delta x|/L$ we have
\begin{align}
\label{eq:leading_order_behavior_midpoint}
{\Delta \bar{S}} &\approx \frac{\Delta t \bar{\xDL}^4}{8} \left( - \frac{ \partial_x^2 \bar{D}}{3} + \frac{(\partial_x \bar{D})^2}{2} \right).
\end{align}
In the midpoint evaluation scheme both $\partial_x^2 \bar{D}$ and $\partial_x \bar{D}$ depend on 
$\Delta x$, because they are always evaluated at the midpoint $x_0 + \Delta x/2$.
Thus, depending on the details of the diffusivity profile,
as $|\Delta x|/L \rightarrow \infty$ 
 the sign of Eq.~\eqref{eq:leading_order_behavior_midpoint} can be either positive or negative,
 or even oscillate.

The short-time propagator in the midpoint evaluation scheme is the usual starting point for
deriving the path-wise entropy production \cite{maes_time-reversal_2003,
seifert_entropy_2005,
seifert_stochastic_2012},
and we now show that the propagator Eq.~\eqref{eq:short_time_propagator_exp_alpha_dt}
with $\alpha = 1/2$ leads to the usual result.

For this, we consider a path $\varphi(t)$ with $t \in [0,t_f]$. 
We discretize the time interval $[0,t_f]$ into $N$ equal subintervals
$[t_i, t_{i+1}]$ with $t_i \equiv t_f \cdot i/N$.
The discretized probability density for the path $\varphi$ is then defined as
\begin{align}
\label{eq:discretized_probability}
P_N[\varphi] &\equiv \prod_{i=0}^{N-1} P(\varphi_{i+1}, t_{i+1} \mid \varphi_i,t_i),
\end{align}
where $\varphi_i \equiv \varphi(t_i)$.
The medium entropy production $\Delta s_{\mathrm{m}}[\varphi]$ of the path $\varphi$ 
is then obtained as the negative log-ratio of the $N \rightarrow \infty$ limit for the
probability density ratio of a pair of forward- and backward path, with the latter defined 
as $\varphi^{b}(t) \equiv \varphi(t_f - t)$ \cite{maes_time-reversal_2003,
seifert_entropy_2005,
seifert_stochastic_2012}. 
Substituting Eq.~\eqref{eq:discretized_probability} for the forward and backward path pair and 
using the midpoint-discretized
propagator 
\cite{seifert_stochastic_2012,
bo_functionals_2019}, we obtain
\begin{align}
\Delta s_{\mathrm{m}}[\varphi]
&\equiv -\lim_{N \rightarrow \infty} \ln \frac{P[\varphi]}{P[{\varphi^{b}}]}
\\ &=
- \lim_{N \rightarrow \infty} \prod_{i=0}^{N-1} \ln \frac{ P(\varphi_{i+1}, t_{i+1}\mid \varphi_i, t_i)}{P(\varphi_{i}, t_{i+1} \mid \varphi_{i+1}, t_i)}
\\ &=
\label{eq:entropy_sum_0}
- \lim_{N \rightarrow \infty}
 \sum_{i=0}^{N-1} \left[
 \Delta \bar{S}(\Delta \varphi_i, \Delta t, \bar{\varphi}_i)
 \right.  \\ & \qquad\qquad\qquad\qquad \left.
 -  
 \Delta \bar{S}(-\Delta \varphi_i, \Delta t, \bar{\varphi}_i)
 \right],
 \nonumber
\end{align}
where $\Delta \varphi_i \equiv \varphi_{i+1} - \varphi_i$, $\bar{\varphi}_i \equiv (\varphi_{i+1}+ \varphi_i)/2$.
From Eq.~\eqref{eq:entropy_sum_0}
 it is apparent that only 
 those terms from $\Delta \bar{S}(\Delta \varphi_i, \Delta t, \bar{\varphi}_i)$ contribute that are odd under 
 the reflection $\Delta \varphi_i \rightarrow - \Delta \varphi_i$.
These are precisely the terms in Eq.~\eqref{eq:delta_bar_S} that are linear in $\Delta x$,
 which are all contained in Eq.~\eqref{eq:delta_bar_S0}.
In particular, all terms of order $\Delta t$ from Eq.~\eqref{eq:delta_bar_S1} 
are even under the reflection $\Delta x \rightarrow - \Delta x$, 
and are hence not relevant for the medium entropy production.
Therefore, the propagator to order $\sqrt{\Delta t} \sim \Delta x$ in the exponent,
which includes the GP, is in fact sufficient to derive the
path-wise entropy production.

After substituting the explicit expression for $\Delta \bar{S}$, we obtain the continuum limit
\begin{align}
\Delta s_{\mathrm{m}}[\varphi]
&= \int_0^T dt \,\dot{\varphi}(t) \frac{ a(\varphi(t)) - (\partial_x D)(\varphi(t))}{D(\varphi(t))} ,
\end{align}
which is the usual formula for the entropy production for a 
multiplicative-noise system \cite{cates_stochastic_2022,
kappler_irreversibility_2020}.

\subsection{Comparison of quadratic-order propagators to
exact solution}

In Sect.~\ref{sec:numerical_example} of the main text we 
compare our perturbative results to
to
an explicit example system.
In the present appendix we extend the comparison to
the quadratic-order propagators from the previous subsections,
namely the NPP Eq.~\eqref{eq:NPP_quadratic},
the PPP with initial-point evaluation Eq.~\eqref{eq:short_time_propagator_exp_dt},
and the PPP with midpoint evaluation, Eq.~\eqref{eq:short_time_propagator_exp_alpha_dt} 
with $\alpha = 1/2$.

\begin{figure*}[ht]
\centering \includegraphics[width=1\textwidth]{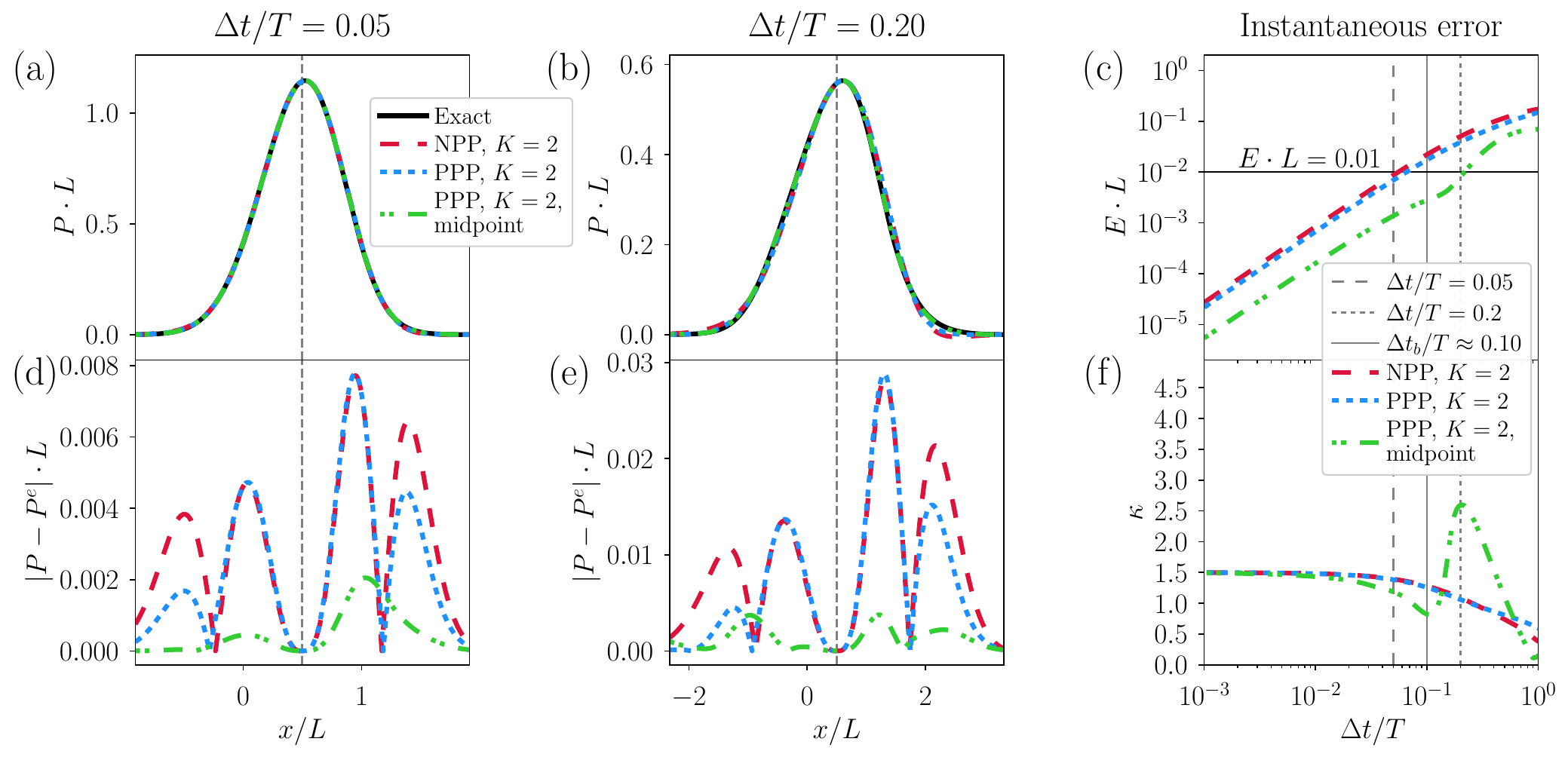} 
\caption{ \label{fig:comparison_np_pp_strato} 
\textbf{Comparison of normalization-preserving propagator (NPP), 
positivity-preserving propagator (PPP), and PPP using a midpoint-evaluation convention.
}
Throughout this figure, we plot data pertaining to the 
 normalization-preserving propagator (NPP) 
 for $K=2$, 
 Eq.~\eqref{eq:NPP_quadratic},
 as blue dotted
 lines.
We furthermore show data based on the 
 positivity-preserving propagator (PPP) for $K=2$,
 Eq.~\eqref{eq:short_time_propagator_exp_dt}
 for $K=2$ 
  as red dashed lines,
 and data obtained using the PPP for $K=2$ with midpoint evaluation,
 Eq.~\eqref{eq:short_time_propagator_exp_alpha_dt} with $\alpha =1/2$,
 as green dash-dotted lines.
For all data we use the the drift and diffusivity from Fig.~\ref{fig:diffusivity_and_drift},
as well as the initial condition $x_0/L = 0.5$.
In subplots (a), (b) we plot the exact solution $P^e$ (black solid line), together
with various approximate propagators;
we show the respective pointwise error Eq.~\eqref{eq:pointwise_error} in subplots (d) and (e).
While subplots (a), (d) correspond to a lagtime $\Delta t/T = 0.05$, for (b), (e) we use $\Delta t/T = 0.2$.
The legend in subplot (a) is valid for subplots (a), (b), (d), (e).
In subplot (c) we show the $L^1$ error Eq.~\eqref{eq:L1_error}
for all approximate propagators considered;
in subplot (f) we plot the corresponding local exponent Eq.~\eqref{eq:running_exponent}.
The vertical broken lines in subplots (c), (f) denote the lagtimes used for subplots (a), (b), (d), (e),
the vertical solid line indicates the breakdown time $\Delta t_b/T \approx 0.10$ as defined in 
Eq.~\eqref{eq:temporal_breakdown_condition}.
The horizontal solid line in subplot (c) indicates the error $E_p \cdot L = 0.01 = 1\%$.
}
\end{figure*}

In Fig.~\ref{fig:comparison_np_pp_strato} (a), (b) we plot 
both 
the exact solution (black solid line) 
and 
the perturbative propagators (broken colored lines)
for (a) $\Delta t/T = 0.05$ and (b) $\Delta t/T = 0.2$.
In Fig.~\ref{fig:comparison_np_pp_strato} (a) 
we observe that for $\Delta t/T = 0.05$, 
all perturbative solutions agree very well with the exact solution on the plotting scales.
We confirm this in Fig.~\ref{fig:comparison_np_pp_strato} (d), where we show
 that the instantaneous
error Eq.~\eqref{eq:pointwise_error} for all approximations is small compared to the
typical function values
of the probability densities. 
The plot also shows that while the error for all approximate propagators is of the same order, 
the NPP with $K=2$ and midpoint evaluation rule leads to a slightly smaller error.

In Fig.~\ref{fig:comparison_np_pp_strato} (b)
 we consider the time increment $\Delta t/T = 0.2$. 
While the PPP with $K=2$ and midpoint evaluation scheme still agrees very well with the
exact solution, 
the other two propagators  deviate visibly from the exact solution for $x/L \approx \pm 2$.
This is also clearly seen in Fig.~\ref{fig:comparison_np_pp_strato} (e),
where we show the pointwise error for $\Delta t/T = 0.2$.

In Fig.~\ref{fig:comparison_np_pp_strato} (c) we show the $L^1$-error Eq.~\eqref{eq:L1_error} 
as a function of $\Delta t$
for all approximate solutions.
While the errors in the NPP and PPP with initial-point evaluation are almost identical in magnitude,
 for $\Delta t \ll \Delta t_b$ these errors are approximately a factor of five larger 
as compared to the error in the PPP with midpoint evaluation.
As the running exponents
Eq.~\eqref{fig:comparison_np_pp_strato} in Fig.~\ref{fig:comparison_np_pp_strato} (f)
show, for all approximate propagators the error
scales as $E \sim \Delta t^{3/2}$ for $\Delta t \ll \Delta t_b$;
this is in agreement with our estimate Eq.~\eqref{eq:ENPP}.

While the PPPs we consider here
 decay to zero at the bounds of the domain we consider here,
we again emphasize that this is not self-evident.
For the diffusivity and drift shown in Fig.~\ref{fig:comparison_np_pp_strato}
and the initial value $x_0/L = 0.5$ we have
$15 (\partial_x D)^2(x_0)/(8D(x_0))T \approx 0.35 > -0.29 \approx (\partial_x^2D)(x_0) T$,
so that according to Eq.~\eqref{eq:condition_BC_exp}
the PPP fulfills the boundary conditions Eq.~\eqref{eq:FP_BC_DL}.
On the other hand, for example for $x_0/L = 1$ it holds that 
$15 (\partial_x D)^2(x_0)/(8D(x_0))T \approx 0.04 < 1.44 \approx (\partial_x^2D)(x_0) T$,
so that the representation Eq.~\eqref{eq:short_time_propagator_exp_dt} diverges
as  $|\Delta x/L| \rightarrow \infty$.

\section{Moments in physical units}
\label{app:moments}

In Sect.~\ref{sec:FTKM}, we discuss the perturbative calculation of the moments $\langle \Delta x^n\rangle$.
In physical units, Eq.~\eqref{eq:n_th_moment_2} yields a perturbation expansion
\begin{align}
\label{eq:delta_xn_physical_units}
\langle \Delta x^n \rangle &= \sum_{k=0}^{\infty} \Delta t^k \langle \Delta x^n \rangle^{(k)}.
\end{align}
Using Eq.~\eqref{eq:xDL_n_k}, we find that 
for $n = 0$, we have $\langle \Delta x^0 \rangle^{(k)} = \delta_{k,0}$.
In the following, we furthermore give the lowest-order terms of Eq.~\eqref{eq:delta_xn_physical_units}
for $n=1, 2, 3$.
In all the expressions, $a$, $D$, and their derivatives are evaluated at $x_0$.

{$n=1$:}
\begin{align}
\langle \Delta x \rangle^{(0)} &= 0,
 \\ 
\langle \Delta x \rangle^{(1)} &= a,
 \\ 
2 \langle \Delta x \rangle^{(2)} &= D a^{(2)}
+a a^{(1)},
 \\ 
6 \langle \Delta x \rangle^{(3)} &= a (a^{(1)})^2
+a^2 a^{(2)}
 %\\ & \quad 
+D^2 a^{(4)}
 \\ & \quad 
 \nonumber
+3D a^{(1)} a^{(2)}
+2D D^{(1)} a^{(3)}
 %\\ & \quad 
+2D a a^{(3)}
 \\ & \quad 
  \nonumber
+D D^{(2)} a^{(2)}
 %\\ & \quad 
+D^{(1)} a a^{(2)},
\end{align}

{$n=2$:}
\begin{align}
\langle \Delta x^2 \rangle^{(0)} &= 0,
 \\ 
\langle \Delta x^2 \rangle^{(1)} &= 2 D,
 \\ 
\langle \Delta x^2 \rangle^{(2)} &= a^2
+D D^{(2)}
+D^{(1)} a
 %\\ & \quad 
+2 D a^{(1)},
 \\ 
3 \langle \Delta x^2 \rangle^{(3)} &= 3 a^2 a^{(1)}
+D (D^{(2)})^2
 %\\ & \quad 
+D^{(2)} a^2
\\ & \quad 
\nonumber
+D^2 D^{(4)}
+4 D (a^{(1)})^2
 %\\ & \quad 
+4 D^2 a^{(3)}
\\ & \quad 
\nonumber
+3D^{(1)} a a^{(1)}
+D^{(1)} D^{(2)} a
 %\\ & \quad 
+2 D D^{(1)} D^{(3)}
\\ & \quad 
+2 D D^{(3)} a
 \nonumber
+4 D D^{(2)} a^{(1)}
+7 D D^{(1)} a^{(2)}
\\ & \quad 
\nonumber
+7 D a a^{(2)},
\end{align}

{$n=3$:}
\begin{align}
\langle \Delta x^3 \rangle^{(0)} &= 0,
 \\ 
\langle \Delta x^3 \rangle^{(1)} &= 0, 
 \\ 
\langle \Delta x^3 \rangle^{(2)} &= 6 D D^{(1)}
+6 D a,
 \\ 
\langle \Delta x^3 \rangle^{(3)} &= a^3
+2 (D^{(1)})^2 a
 %\\ & \quad 
+3 D^{(1)} a^2
 \\ & \quad 
 \nonumber
+4 D^2 D^{(3)}
 %\\ & \quad 
+7 D^2 a^{(2)}
 %\\ & \quad 
+7 D D^{(2)} a
 \\ & \quad 
  \nonumber
+8 D D^{(1)} D^{(2)}
 %\\ & \quad 
+9 D a a^{(1)}
 %\\ & \quad 
+10 D D^{(1)} a^{(1)}.
\end{align}
These expressions, as well as higher order terms up to $\langle \Delta x^n \rangle^{(k)}$
for $n = 0, 1, 2, 3, 4$ and $k = 0, 1, ..., 4$,
are given as symbolic expressions in the python module PySTFP \cite{PySTFP}.

\section{Perturbative entropy production}
\label{app:entropy}

Following the stochastic thermodynamics literature \cite{seifert_entropy_2005,
seifert_stochastic_2012} 
we define the non-equilibrium Gibbs entropy
\begin{align}
\label{eq:S_gibbs_app}
S(t) &\equiv -\int_{-\infty}^{\infty}dx\,P(x,t) \ln \left[ P(x,t) L \right],
\end{align}
where here $P(x,t) \equiv P(x,t\mid x_0,t_0)$, 
and where in the logarithm we multiply $P$
by the length scale $L$ in order to render the argument of the logarithm dimensionless.
By taking the time derivative of this entropy, and using the FP Eq.~\eqref{eq:fokker_planck}, 
we obtain \cite{seifert_entropy_2005,
seifert_stochastic_2012}
\begin{align}
\dot{S} (t) &= \dot{S}^{\mathrm{tot}}(t) - \dot{S}^{\mathrm{m}}(t),
\end{align}
with the total- and medium entropy production
\begin{align}
\label{eq:S_tot_app}
\dot{S}^{\mathrm{tot}}(t) 
&\equiv
\int_{-\infty}^{\infty}dx\,\frac{ j(x,t)^2}{D(x)P(x,t)},
\\
\label{eq:S_m_app}
\dot{S}^{\mathrm{m}}(t) 
&\equiv
\int_{-\infty}^{\infty}dx\,\frac{ j(x,t)}{D(x)}\left[
a(x) - (\partial_x D)(x)
 \right],
\end{align}
with the standard FP probability flux
\begin{align}
\label{eq:j_def_app}
j &\equiv aP - \partial_x (D P).
\end{align}

To evaluate
Eqs.~\eqref{eq:S_gibbs_app},
 \eqref{eq:S_tot_app},
  \eqref{eq:S_m_app}
  perturbatively,
   we first 
rewrite the expressions 
in dimensionless form using 
Eqs.~(\ref{eq:def_epsDL}-\ref{eq:def_DDL}).
This yields
\begin{align}
\label{eq:tilde_S}
\tilde{S}(\tDL) &\equiv
{S}(t)
=
- \int_{-\infty}^{\infty} d\xDL\,\PDL(\xDL,\tDL) \ln \left[ \frac{\PDL(\xDL,\tDL) }{\epsDL(\tDL)}\right],
\\
\label{eq:tilde_S_tot}
\dot{\tilde{S}}^{\mathrm{tot}}(\tDL) &\equiv \tD \dot{S}^{\mathrm{tot}}(t) 
=\epsDL^2 \int_{-\infty}^{\infty}d\xDL
\frac{
\jDL(\xDL,\tDL)^2
}{
\DDL(\xDL,\tDL) \PDL(\xDL,\tDL)
},
\\
\label{eq:tilde_S_m}
\dot{\tilde{S}}^{\mathrm{m}}(\tDL) &\equiv \tD \dot{{S}}^{\mathrm{m}}(t)
= 
\int_{-\infty}^{\infty}d\xDL
\frac{
\jDL(\xDL,\tDL)
}{
\DDL(\xDL,\tDL)
}
\left[ 
\epsDL \aDL - (\partial_{\xDL} \DDL) 
\right],
\end{align}
with the dimensionless probability flux
\begin{align}
\jDL &\equiv \tD j \equiv \frac{1}{\epsDL^2}\left[  \epsDL \aDL \PDL - 
 \partial_{\xDL} (\DDL \PDL) \right].
\end{align}
We now discuss the perturbative evaluation of Eqs.~\eqref{eq:tilde_S},
\eqref{eq:tilde_S_tot},
\eqref{eq:tilde_S_m}.

First, to evaluate Eq.~\eqref{eq:tilde_S} perturbatively,
we substitute the power series expansion Eq.~\eqref{eq:perturbative_solution_rewritten_1} 
for $\PDL$ into the
 logarithm and use the standard rules for manipulating logarithms, as well as the 
 normalization of the perturbative propagator, to derive
\begin{align}
\label{eq:S_Gibbs_rewritten_DL}
\tilde{S}(\tDL) 
=&
 \frac{1}{2}\ln(2 \pi \epsDL^2) 
+
\frac{1}{2}
 \int_{-\infty}^{\infty} d\xDL\,\PDL(\xDL,\tDL)   \xDL^2
\\ &\nonumber
- 
 \int_{-\infty}^{\infty} d\xDL\,\PDL(\xDL,\tDL)  \ln \left[1 + \sum_{k=1}^{\infty}\epsDL(\tDL)^k \Qindex{k}(\xDL,\tDL)\right].
\end{align}
Upon expanding the logarithm in the second line 
in powers of $\epsDL$, and substituting 
Eq.~\eqref{eq:perturbative_solution_rewritten_1} into the integrands,
both integrands in Eq.~\eqref{eq:S_Gibbs_rewritten_DL}
 become sums over terms that are each a product of a polynomial in $\xDL$ with
a Gaussian.
These integrals are easily evaluated,  so that to leading order we obtain
\begin{align}
\label{eq:S_gibbs_power_series_DL}
\tilde{S}
&=
  \frac{1}{2}\left[ 1  + \ln(  2 \pi \epsDL^2 )\right]
  \\ & \qquad 
  + \frac{\epsDL^2 }{16}
  \left[ 2 \mathcal{A}_0 \mathcal{D}_1
  + 4 \mathcal{A}_1 
  - 3 \mathcal{D}_1^2 
  + 4 \mathcal{D}_2
  \right]
  + \mathcal{O}(\epsDL^4).
  \nonumber
\end{align}
In the python module PySTFP \cite{PySTFP} we provide the power series coefficients
up to order $\epsDL^8 \sim \Delta t^4$ in symbolic form.

We now evaluate the medium entropy production rate Eq.~\eqref{eq:tilde_S_m} perturbatively.
For this we first use integration by parts to rewrite the integral as
\begin{align}
\label{eq:S_dot_m_dimensionless_rewritten}
\dot{\tilde{S}}^{\mathrm{m}}(\tDL) =&
 \int_{-\infty}^{\infty} d\xDL\,\PDL(\xDL,\tDL)
 \\&\times \nonumber
 \left[
  \frac{1}{\DDL}
  \left(  \aDL - \frac{1}{\epsDL}\partial_{\xDL} \DDL \right)^2
  + \frac{1}{\epsDL}\partial_{\xDL} 
    \left(  \aDL -\frac{1}{\epsDL} \partial_{\xDL} \DDL \right)
  \right],
\end{align}
where in our notation we suppress the arguments of $\aDL(\xDL,\tDL)$, 
$\DDL(\xDL,\tDL)$, $\epsDL(\tDL)$.
Equation \eqref{eq:S_dot_m_dimensionless_rewritten} has the form of an 
expectation value integral, and by substituting the perturbation series 
Eq.~\eqref{eq:perturbative_solution_rewritten_1}, 
\eqref{eq:aDL_power_series}, 
\eqref{eq:DDL_power_series} 
in the integral, we evaluate it perturbatively to any desired order.
We note that despite the explicit appearance of a factor $1/\epsDL^2$ in 
Eq.~\eqref{eq:S_dot_m_dimensionless_rewritten}, the leading order
is actually $\epsDL^0$.
This follows
by substituting the perturbation series for $\aDL$, $\DDL$, and evaluating the
spatial derivatives that appear in the second factor in 
Eq.~\eqref{eq:S_dot_m_dimensionless_rewritten};
consequently, we conclude that $\dot{\tilde{S}}^{\mathrm{m}} = \mathcal{O}(\epsDL^0)$.
More explicitly, to leading order in $\epsDL$ we obtain
\begin{align}
\label{eq:S_m_power_series_DL}
\dot{\tilde{S}}^{\mathrm{m}} 
&= 
\left( \mathcal{A}_0 - \mathcal{D}_1\right)^2  + \mathcal{A}_1 - 2 \mathcal{D}_2
+ \mathcal{O}(\epsDL^2).
\end{align}
In the python module PySTFP \cite{PySTFP} we provide the power series coefficients
up to order $\epsDL^8 \sim \Delta t^4$ in symbolic form.

By contrast, since the integrand in the
 total entropy production rate Eq.~\eqref{eq:tilde_S_tot} is quadratic
 in $\PDL$, the expression cannot be rewritten 
in the form of an expectation value of a $\PDL$-independent function.
We can evaluate the integral nonetheless perturbatively, by substituting the power series
expansions of $\PDL$, $\aDL$, $\DDL$, sorting the resulting integrand by powers of $\epsDL$,
and evaluating the integrals (which 
are products of polynomials and a Gaussian)
up to the desired order.
The resulting leading order contributions are
\begin{align}
\dot{\tilde{S}}^{\mathrm{tot}} =&
\frac{1}{\epsDL^2} + 
\frac{1}{8} \left( 
8\Aindex{0}
- 14 \Aindex{0} \Dindex{1}
+ 12 \Aindex{1}
+ 5 \Dindex{1}^2
- 12 \Dindex{2}
\right)
\nonumber
\\& + \mathcal{O}(\epsDL^2).
\label{eq:S_tot_power_series_DL}
\end{align}
In the python module PySTFP \cite{PySTFP} we provide the power series coefficients
up to order $\epsDL^6 \sim \Delta t^3$ in symbolic form.

Recasting the power series Eqs.~\eqref{eq:S_gibbs_power_series_DL}, 
\eqref{eq:S_m_power_series_DL}, 
\eqref{eq:S_tot_power_series_DL}
 in physical dimensions we then obtain
\begin{align}
\label{eq:S_gibbs_physical_dimensions_app}
{{S}}
&= 
  \frac{1}{2}\left[ 1  + \ln\left(  \frac{4 \pi  D \Delta t}{L^2} \right)\right]
  \\\nonumber
  & \quad + 
  \frac{\Delta t}{8 D} \left[ 
  2 a \partial_x D 
  + 4 (\partial_x a )D
  + 2 (\partial_x^2 D)D
  - 3 (\partial_x D)^2
  \right]
  \\ \nonumber & \quad 
  + \mathcal{O}(\Delta t^2),
\\
\dot{{S}}^{\mathrm{m}}
&= 
 %\dot{{S}}^{\mathrm{m},(0)}(t) =&
 \frac{1}{D}\left( a - \partial_x D\right)^2  + \partial_x a - \partial_x^2 D
+   \mathcal{O}(\Delta t),
\\
\dot{S}^{\mathrm{tot}} &=
\frac{1}{2\Delta t}
+ \frac{1}{8D}
\left[ 8 a^2 
- 14 a \partial_x D
+ 12 (\partial_x a) D
\right.
\\ &\qquad\qquad\qquad \left. \nonumber
- 6 D \partial_x^2 D
+ 5 (\partial_x D)^2
\right]
+ \mathcal{O}(\Delta t),
\end{align}
where $a$, $D$, and their derivatives are evaluated at the initial position $x_0$.
We note that Eq.~\eqref{eq:S_gibbs_physical_dimensions_app} explicitly depends on $L$, 
which is because in Eq.~\eqref{eq:S_gibbs_app} we include a factor $L$ to render the argument
of the logarithm dimensionless.

\section{Medium entropy production as rate of change of potential energy}
\label{eq:app_medium_potential}

We now discuss how Eq.~\eqref{eq:S_m_app}
can be reformulated in terms of a potential energy for systems where the 
latter is defined \cite{seifert_stochastic_2019,
sekimoto_stochastic_2010}.

We consider a system where a zero-flux steady state $P_{\mathrm{steady}}$ exists, which
by definition fulfils
\begin{align}
\partial_x (D P_{\mathrm{steady}}) - a P_{\mathrm{steady}} &= 0
\end{align}
everywhere.
Integrating this equation it follows that
\begin{align}
P_{\mathrm{steady}}(x) &= \frac{ \mathcal{N}(x_{\mathrm{ref}})}{D(x)} \exp\left[ 
 \int_{x_{\mathrm{ref}}}^{x}dx'
\frac{a(x')}{D(x')} 
\right]
\end{align}
where $x_{\mathrm{ref}}$ is arbitrary and $\mathcal{N}(x_{\mathrm{ref}})$ is 
a normalization constant so that $P_{\mathrm{steady}}$ is a properly normalized probability
density.

Via Boltzmann inversion, we obtain the corresponding potential $U(x)$ as
\begin{align}
U(x) =& 
- \ln \left[ P_{\mathrm{steady}}(x)L \right]
\\
=& 
\ln\left[ \frac{ D(x)}{D_0}\right]
- 
 \int_{x_{\mathrm{ref}}}^{x}dx'
\frac{a(x')}{D(x')} 
+ U_0,
\label{eq:def_U_1_app}
\end{align}
where $U_0 = - \ln[L \mathcal{N}(x_{\mathrm{ref}})/ D_0]$ is independent of $x$.
Note that our definition of $U$ makes the potential a dimensionless quantity, 
which one can think of the as the potential being expressed in units of the thermal energy 
(what we write $U$ here is sometimes denoted as $\beta U$).

Taking the derivative of Eq.~\eqref{eq:def_U_1_app} with respect to $x$, we obtain
\begin{align}
\partial_x U =&
-\frac{1}{D(x)}\left[
a(x) - (\partial_x D)(x)
\right]
\end{align}
We therefore have 
\begin{align}
\dot{S}^{\mathrm{m}}(t) 
\equiv&
\int_{-\infty}^{\infty}dx\,\frac{ j(x,t)}{D(x)}\left[
a(x) - (\partial_x D)(x)
 \right]
 \label{eq:Sm_rewrite_app_0}
 \\ =&
- \int_{-\infty}^{\infty}dx\,j(x,t) 
(\partial_x U)(x)
\label{eq:Sm_rewrite_app_1}
 \\ =&
 \int_{-\infty}^{\infty}dx\,(\partial_x j)(x,t) 
(U)(x)
\label{eq:Sm_rewrite_app_2}
 \\ =&
-\int_{-\infty}^{\infty}dx\,(\partial_t P)(x,t) 
U(x)
\label{eq:Sm_rewrite_app_3}
\\ = &
- \partial_t \langle  U \rangle,
\label{eq:Sm_rewrite_app_4}
\end{align}
where at Eq.~\eqref{eq:Sm_rewrite_app_1}
we use integration by parts (and that vanishing-flux boundary conditions at $x = \pm \infty$ for $j$),
at Eq.~\eqref{eq:Sm_rewrite_app_3} we 
use the FP Eq.~\eqref{eq:fokker_planck}
and at Eq.~\eqref{eq:Sm_rewrite_app_4}
we use that $U$ is time-independent.

\section{Analytical solution for multiplicative-noise example system}
\label{app:example_system}

For our numerical examples,
we now construct a nontrivial 
one-dimensional multiplicative-noise system
 with a known analytical solution.
 For that we first consider 
  the free-diffusion
Fokker-Planck equation
\begin{align}
\label{eq:fokker_planck_y_coordinates}
\partial_t P_Y  &= D_0 \partial_y^2 P_Y
\end{align}
in a coordinate system $(y,t)$,
and with a constant diffusivity $D_0$.
With delta-peak initial condition at $y_0$, and vanishing boundary conditions at $|y| \rightarrow \infty$,
Eq.~\eqref{eq:fokker_planck_y_coordinates}
is solved by
\begin{align}
\label{eq:fokker_planck_y_coordinates_P}
P_Y(y,t \mid y_0, t_0) &= \frac{ 1}{ \sqrt{ 4 \pi D_0 \Delta t } } \exp\left[ - \frac{ (y-y_0)^2}{4 D_0 \Delta t}\right],
\end{align}
where $\Delta t = t - t_0$.

We define a coordinate transformation $x \equiv \Phi(y)$,
and rewrite the Fokker-Planck Eq.~\eqref{eq:fokker_planck_y_coordinates}
with respect to the new coordinate $x$.
This leads to \cite{risken_fokker-planck_1996}
\begin{align}
\label{eq:fokker_planck_x_coordinates}
\partial_t P &= - \partial_x \left( a P \right) + \partial_x^2 (DP),
\end{align}
where 
\begin{align}
\label{eq:fokker_planck_x_coordinates_a}
a(x) &= D_0 \left.(\partial_y^2 \Phi)\right|_{y = \Phi^{-1}(x)},
\\
\label{eq:fokker_planck_x_coordinates_D}
D(x) &= D_0 \left[ \left.\left(\partial_y \Phi\right)\right|_{y = \Phi^{-1}(x)}\right]^2.
\end{align}
The solution to Eq.~\eqref{eq:fokker_planck_x_coordinates}, 
\eqref{eq:fokker_planck_x_coordinates_a}, 
\eqref{eq:fokker_planck_x_coordinates_D}
 with delta-peak initial condition at $x_0$
is then given by
\begin{align}
\label{eq:fokker_planck_x_coordinates_P}
P(x,t \mid x_0, t_0) &=\left.\left[ \left(\dfrac{ d \Phi}{d y}\right)^{-1} 
P_Y\left( y , t\mid y_0 ,t_0\right)\right]\right|_{{y = \Phi^{-1}(x)}},
\end{align}
where $y_0 = \Phi^{-1}(x_0)$ and $P_Y$ is given by Eq.~\eqref{eq:fokker_planck_y_coordinates_P}.

For the explicit example system considered in the main text we introduce a length scale $L$
and a time scale $T$, and set
$D_0 = L^2/T$.
As coordinate transformation we use
\begin{equation}
\label{eq:app_coordinate_transformation}
\Phi(y) \equiv 0.35 \pi y + 0.025 L \sin\left(\pi \frac{y}{L} \right),
\end{equation}
so that it follows from Eqs.~\eqref{eq:fokker_planck_x_coordinates_a}, 
\eqref{eq:fokker_planck_x_coordinates_D}, 
that
\begin{align}
a(x) &= -\frac{L}{T} 0.025 \pi^2 \sin\left( \pi \frac{\Phi^{-1}(x)}{ L}\right),
\\
D(x) &= \frac{L^2}{T}\pi \left[ 0.35  + 0.025  \cos\left( \pi \frac{\Phi^{-1}(x)}{ L}\right)\right]^2,
\\
P(x,t \mid x_0, t_0) &=
\frac{1}{\pi} \left[ 0.35  + 0.025  \cos\left( \pi \frac{\Phi^{-1}(x)}{ L}\right)\right]^{-1}
\\
& \quad 
\times 
\frac{ 1}{ \sqrt{ 4 \pi D_0 \Delta t } } \exp\left[ - \frac{ (\Phi^{-1}(x)-\Phi^{-1}(x_0))^2}{4 D_0 \Delta t}\right].
\nonumber
\end{align}
To evaluate $\Phi^{-1}(x)$ numerically  
we use that for any $x$, finding $y$ such that $y = \Phi^{-1}(x)$ is equivalent to
finding the root of the function $g(y) \equiv \Phi(y) - x$.
To do this in our numerical implementation we use the function root\_scalar
from scipy.optimize \cite{2020SciPy-NMeth}.

For our perturbative solution we need to calculate the derivatives 
of $a$, $D$, $P$, at $x_0$. 
For this, we note that Eqs.~\eqref{eq:fokker_planck_x_coordinates_a}, 
\eqref{eq:fokker_planck_x_coordinates_D},  
 \eqref{eq:fokker_planck_x_coordinates_P}
are all of the form
\begin{align}
\label{eq:f_x_app}
f(x) &= g(\Phi^{-1}(x)),
\end{align}
for explicitly given $g$. 
For example, for Eq.~\eqref{eq:fokker_planck_x_coordinates_D} we have
$f = D$ and $g = D_0 (\partial_y \Phi)^2$.
We furthermore note that
from Eq.~\eqref{eq:app_coordinate_transformation}, 
we can straightforwardly evaluate the derivatives of $\Phi$ with respect to $y$.
We thus seek a formula for the derivatives of $f$,
in terms of the derivatives of $g$ and $\Phi$.
To obtain such a formula, we note that Eq.~\eqref{eq:f_x_app} is equivalent to
\begin{align}
\label{eq:f_x_app_2}
f(\Phi(y)) &= g(y).
\end{align}
By taking $n$ derivatives of this equation with respect to $y$,
we obtain an analytical formula for $(\partial_x^n f)(\Phi(y))$
with the need to know the explicit form of $\Phi^{-1}$.
We then evaluate the derivative at $x_0$ by substituting $y$ 
with a numerically calculated $y_0 \equiv \Phi^{-1}(x_0)$.
We use this procedure to calculate all analytical derivatives for $a$, $D$, $P$.
The code for all evaluations of the analytical solution, 
and the derivatives of drift, diffusivity, and transition density, are provided in the python
module PySTFP \cite{PySTFP}.

\end{document}